%% file: matchbox.tex
\documentclass[epj]{svjour}
\usepackage{amsmath,amssymb,amsfonts,graphicx,cite,algorithmic,algorithm,multirow,array}
\graphicspath{{graphics/}}
\makeatletter
\ifx\input@path\@undefined
\def\input@path{{graphics/}}
\else
\g@addto@macro\input@path{{graphics/}}
\fi
\makeatother
\newcommand{\program}[1]{\textsf{#1}}

\newcommand{\thepeg}{\textsc{ThePEG}}
\newcommand{\exsample}{\textsc{ExSample}}
\newcommand{\matchbox}{\textsc{Matchbox}}
\newcommand{\rivet}{\textsc{Rivet}}
\newcommand{\professor}{\textsc{Professor}}
\newcommand{\herwig}{\textsc{Herwig}}
\newcommand{\hpp}{\textsc{Herwig++}}

\newcommand{\sherpa}{\textsc{sherpa}}
\newcommand{\powheg}{\textsc{powheg}}

\preprint{DESY 11-162\\KA-TP-24-2011\\HERWIG-11-01\\MCnet-11-24}

\title{Dipole Showers and Automated NLO Matching in Herwig++}

\author{Simon Pl\"atzer\inst{1} \and Stefan Gieseke\inst{2}}

\institute{DESY, Notkestrasse 85, D-22607 Hamburg, Germany\and%
Institut f\"ur Theoretische Physik, KIT, D-76128 Karlsruhe, Germany}

\date{\today}

\abstract{We report on the implementation of a coherent dipole
  shower algorithm along with an
  automated implementation for dipole subtraction and for performing
  \powheg{}- and MC@NLO-type matching to next-to-leading order (NLO)
  calculations.  Both programs are implemented as add-on modules to
  the event generator \hpp{}. A
  preliminary tune of parameters to data acquired at LEP, HERA and
  Drell-Yan pair production at the Tevatron has been performed, and we
  find an overall very good description which is slightly improved by
  the NLO matching.   }

\begin{document}

\maketitle


\section{Introduction}
\label{sections:introduction}

Many physics analyses at the Large Hadron Collider (LHC) are nowadays
based on Monte Carlo simulations
\cite{Bahr:2008pv,Corcella:2002jc,Sjostrand:2007gs,Sjostrand:2006za,Gleisberg:2008ta},
e.g.\ for acceptance determination or even for background subtraction.
With the high precision aimed for in many analyses it is mandatory to
provide many of the simulations with the highest possible theoretical
accuracy.  For most processes this is now next-to-leading order (NLO) in
the perturbative expansion of Quantum Chromodynamics (QCD).  During the
last decade, enormous progress was made in the development of techniques
to match NLO calculations on the one hand and to merge multiple jet tree
level matrix elements on the other hand with parton shower algorithms.

First attempts to improve parton shower emission patterns with the
information from the full matrix element for the hardest gluon
emission were made with so-called matrix element corrections
\cite{Seymour:1994df,Norrbin:2000uu}, that have long been implemented in
the standard event generators.  The next big improvement was made when
matrix elements for multiple hard emissions were merged with parton
shower algorithms, first for $e^+e^-$ annihilation processes
\cite{Catani:2001cc,Lonnblad:2001iq} and then also for hadronic
collisions \cite{Krauss:2002up}.  An alternative approach was proposed
in \cite{Hoche:2006ph}, where different implementations have been
systematically compared as well.  The experience that was made with
these algorithms over the last years \cite{Lavesson:2007uu} has lead to
further improvements \cite{Hoeche:2009rj,Hamilton:2009ne} such that now
the systematic uncertainties due to e.g.\ matching scale dependence have
been significantly reduced.

Matching to NLO matrix elements has been initiated first with a phase
space slicing method \cite{Potter:2000an,Potter:2001ej,Dobbs:2001dq}.
A more systematic matching has then been introduced by Frixione and
Webber in the MC@NLO approach \cite{Frixione:2002ik}.  This approach
has then been generalised to include massive partons
\cite{Frixione:2003ei}.  Many processes have been included in the
meantime \cite{Frixione:2005vw,Frixione:2007zp,Frixione:2008yi}.  As
the algorithm depends on subtraction terms for a specific parton
shower implementation, the first versions of MC@NLO have been tailored
to work with \herwig{} only.  Now, it also works with \hpp{}, i.e.\ as
the subtraction scheme has been generalised towards the \hpp{} parton
shower implementation, all processes available in the MC@NLO package
can also be showered with \hpp{} to achieve formal accuracy at NLO
\cite{Frixione:2010ra}.

As the matching of NLO matrix elements and parton shower algorithms
takes place perturbatively to the specified order, i.e.\ the
next-to-leading order, there is formally an ambiguity left that can be
used to devise alternative matching schemes.  One such scheme has been
proposed by Nason \cite{Nason:2004rx} and now goes under the name
\powheg{}.  The guiding principle of this algorithm is to allow for a
matching algorithm that does not introduce events with negative
weight, as the MC@NLO prescription does.  This approach has also been
very successfully established during the last years and implemented as
a separate program package \cite{Frixione:2007nu}.  Many processes are
available in this program package
\cite{Nason:2006hfa,Frixione:2007nw,Alioli:2008gx,Alioli:2008tz,Alioli:2010xa}.
However, the method itself is also used by other groups to match NLO
calculations with parton showers within a given shower package.  Many
processes are available with \hpp{}
\cite{Hamilton:2008pd,Hamilton:2009za,Hamilton:2010mb,D'Errico:2011um,D'Errico:2011sd}
or \sherpa{} \cite{Hoche:2010pf}.  The internal implementations
benefit from the inclusion of truncated showers (see below).

On the parton shower side, a number of new parton shower algorithms have
been developed during the last years, partly together with the rewrite
of old generators \cite{Gieseke:2003rz,Sjostrand:2004ef}.  Many new
developments have addressed the idea of implementing a shower that is
directly related to the subtraction terms commonly used in NLO
calculations.  This led to the implementation of parton showers with
splitting kernels based on the Catani--Seymour subtraction scheme
\cite{Catani:1996vz,Catani:2002hc} for NLO calculations
\cite{Schumann:2007mg,Dinsdale:2007mf}, which was proposed in
\cite{Nagy:2005aa}.  Similar ideas are followed with other subtraction
schemes as e.g.\ in the \textsc{vincia} shower \cite{Giele:2007di} where
QCD antenna subtraction terms are facilitated.

With more and more NLO calculations being matched one-by-one the
question arises whether this step can be automated.  In fact, the
\powheg{} method is already a first step into this direction, as the
method as such is independent of the showering algorithm.  In
particular, no specific subtraction terms or the like are needed in
order to match a given NLO calculation to any shower.  There are
subtleties on the shower side, though.  The \powheg{} method guarantees
to give the hardest emission within the parton evolution and ensures
that this is generated according to the phase space weighting of the NLO
matrix element.  However, if the shower does not evolve in the same
hardness measure as the \powheg{} algorithm, one has to introduce
so-called truncated showers.  This has been discussed already in early
\textsc{powheg} implementations \cite{LatundeDada:2006gx} and is now
part of \hpp{} \cite{Hamilton:2009ne} and \sherpa{}
\cite{Hoeche:2009rj}.

Many NLO calculations are available as ready-to-use computer codes that
often come as packages that include a number of processes at NLO
already.  Most of these codes use the Catani--Seymour subtraction method
to regularise infrared divergences.  More recently, also the complete
automation of NLO calculations has been discussed with first tools
readily available \cite{vanHameren:2009dr,Hirschi:2011pa}, based on the
approach \cite{Ossola:2007ax}.  Some more calculations are already based
on a fully automated tool chain
\cite{Berger:2009zg,Berger:2010zx,Giele:2008bc,Mastrolia:2010nb,Heinrich:2010ax}. Part
of this progress relies on the automatic generation of Catani--Seymour
subtraction terms
\cite{Gleisberg:2007md,Hasegawa:2009tx,Frederix:2010cj} or FKS
subtraction terms \cite{Frederix:2009yq}.  The latest developments unify
the matching of multiple tree--level emissions and the matching of NLO
corrections to the Born level \cite{Hamilton:2010wh,Hoche:2010kg}.

In this paper we introduce an implementation of a parton shower based
on the Catani--Seymour subtraction terms, similar to the showers
introduced in \cite{Schumann:2007mg,Dinsdale:2007mf}.  The goal of the
implementation is to provide a framework for an automatic matching of
NLO computations to a parton shower.  The use of the subtraction terms
is highly beneficial as the MC@NLO like matching, that is based on a
subtraction of the parton shower contribution to the NLO observable
becomes trivial.  Together with a framework to handle \textsc{powheg}
like matching we will have the possibility to check systematics within
a single implementation.  By using a shower based framework we may
directly make use of truncated showers in order to minimise systematic
uncertainties inherent to the matching formalism.  As a first step in
this programme we present the shower implementation, which is embedded
as a module in the \hpp{} event generator.  In addition we present NLO
matchings to the basic QCD processes.

The paper is organised as follows.  In Sec.~\ref{sections:dipoleshower}
we introduce the dipole shower in detail.  Sec.~\ref{sections:matchbox}
introduces the implementation of an automatic matching with this parton
shower, that we call \matchbox{}. In Secs.~\ref{sections:lep},
\ref{sections:hera} and \ref{sections:tev} we present comparisons to
data from LEP, HERA and the Tevatron, respectively.  

\section{Dipole Showers}
\label{sections:dipoleshower}

The dipole shower algorithm outlined in \cite{Platzer:2009jq} has been
implemented as an add-on module to \hpp{},
\cite{Bahr:2008pv}. In this section we briefly review its properties
and give a full description of the implementation.

The authors have shown that parton showers based on Catani-Seymour
subtraction kernels \cite{Catani:1996vz} correctly reproduce the
Sudakov anomalous dimensions and properly include effects of soft
gluon coherence, upon using an ordering of emissions in transverse
momenta as defined by the emitting dipoles. The simple inversion of
the kinematic parametrisation used in the context of NLO subtraction,
however, does not resemble a physical picture for initial state
radiation. An alternative has been suggested and implemented in
the simulation presented here.

\subsection{Starting the Shower}

The dipole shower starts evolving off a hard sub process, which is
assigned colour flow information in the large-$N_c$ limit. This colour
flow information is used to first sort all coloured partons attached
to the hard sub process into colour singlets.  Practically, this is
done by making use of the fact that a colour singlet is `simply
connected' in the sense of its colour flow topology: Any parton $i$ in
a colour singlet can be reached from a parton $j$ in the same singlet
by just following colour lines and changing from a colour to an
anti-colour line at an external gluon.  Each colour singlet is now an
independently evolving entity, and can only split into two colour
singlets in the presence of a $g\to q\bar{q}$ splitting.  In the next
step, the partons in each singlet are sorted such that colour
connected partons are located at neighbouring positions, when
representing the singlet group of partons as a sequence. Note that
these sequences may be open or closed: We will call a sequence open,
or non-circular, if there exists a circular permutation of the
elements in it such that the partons at the first and last position
are not colour connected.  Conversely, if there does not exist such a
permutation, the sequence is called circular or closed. The possible
sequences are depicted in Fig.~\ref{figures:chainsplittings}. Once
this sorting has been accomplished, we will refer to these singlet
sequences as {\it dipole chains}: each pair of subsequent partons in a
singlet sequence forms a dipole, which may radiate. For each parton in
each dipole, a hard scale is then determined as defined in
\cite{Platzer:2009jq}.

\subsection{Evolution of the Parton Ensemble}

The main shower algorithm acts on a set of dipole chains, and proceeds
as long as this set is not empty. Dipole chains are removed from the
list, if they stopped evolving, {\it i.e.} if there was no splitting
selected with a $p_\perp^2$ above the shower's infrared cutoff
$\mu_{IR}^2$.  The first entry in the set of dipole chains is taken to
be the current chain. For each dipole $(i,j)$ in the current chain
(with both possible emitter--spectator assignments, {\it i.e.} also
considering $(j,i)$ along with $(i,j)$), any possible splitting
$(i,j)\to (i',k,j)$ is considered to compete with all other possible
splittings of the chain.  For any such splitting, given a hard scale
$p_\perp^2$ associated to the emitter under consideration, a scale
$q_\perp^2$ is selected with probability given by the Sudakov form
factor
\begin{multline}
\Delta_{(i,j)\to (i',k,j)}(q_\perp^2,p_\perp^2) = \\
\exp\left(-\int_{q_\perp^2}^{p_\perp^2}{\rm d}q^2\int_{z_-(q^2)}^{z_+(q^2)}{\rm d}z P_{(i,j)\to (i',k,j)}(q^2,z)\right) \ ,
\end{multline}
where $P_{(i,j)\to (i',k,j)}(q^2,z)$ is the appropriate splitting probability
as defined in \cite{Platzer:2009jq}, using the respective dipole
splitting function $V_{i',k;j}$.

The splitting with the largest selected value of $q_\perp^2$ is then
chosen to be the one to happen, except the largest $q_\perp^2$ turned
out to be below the infrared cutoff. In this case the current chain is
removed from the set of dipole chains, inserted into the event record
and the algorithm proceeds with the next chain.  The momentum fraction
$z$ is chosen to be distributed according to ${\rm d}P_{(i,j)\to
  (i',k,j)}(q_\perp^2,z)$. Since for now we use azimuthally averaged
splitting kernels, the azimuthal orientation of the transverse
momentum is chosen to be distributed flat. The momenta of the
splitting products and the spectator after emission are then
calculated as specified in \cite{Platzer:2009jq}.

\begin{figure}
\begin{center}

\begin{tabular}{m{1.5cm}m{0.5cm}m{1.5cm}m{0.5cm}m{1.5cm}}
\includegraphics[scale=0.3]{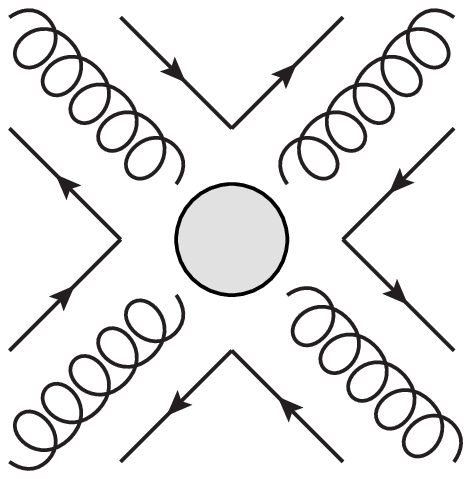}\hspace*{1cm} & {\huge $\to$} &
\includegraphics[scale=0.3]{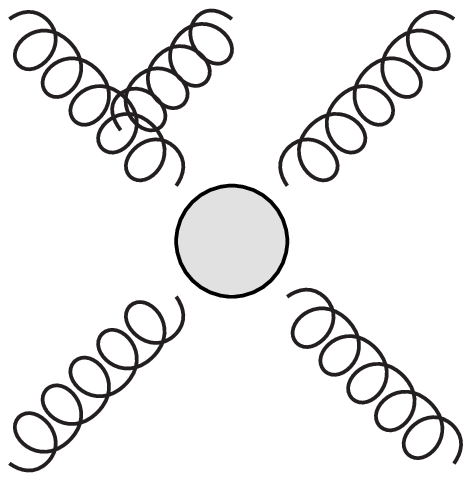}\hspace*{1cm} & {\huge $\to$} &
\includegraphics[scale=0.3]{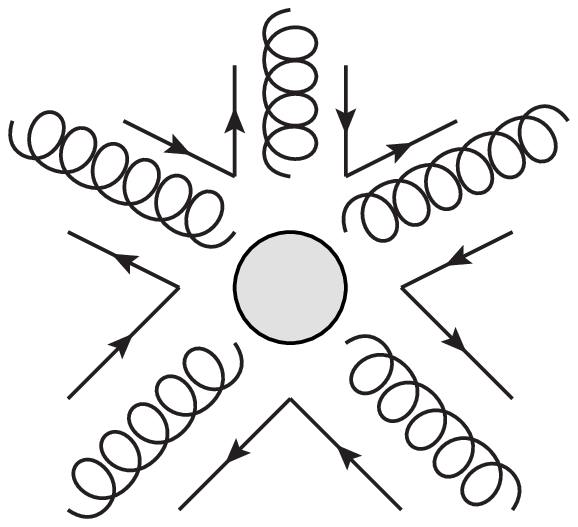}
\end{tabular}

Gluon emission off a circular chain.\\ The chain stays circular.

\begin{tabular}{m{1.5cm}m{0.5cm}m{1.5cm}m{0.5cm}m{1.5cm}}
\includegraphics[scale=0.3]{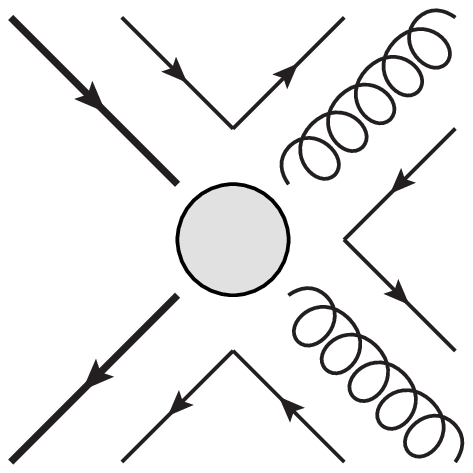}\hspace*{1cm} & {\huge $\to$} &
\includegraphics[scale=0.3]{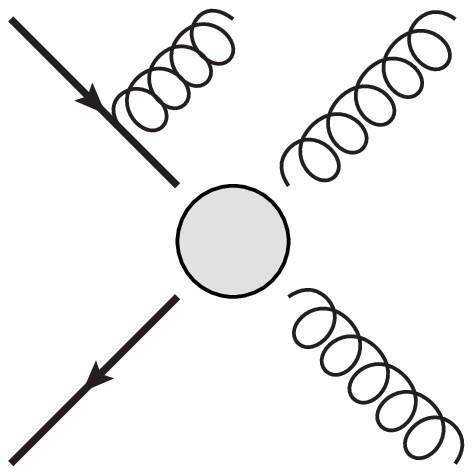}\hspace*{1cm} & {\huge $\to$} &
\includegraphics[scale=0.3]{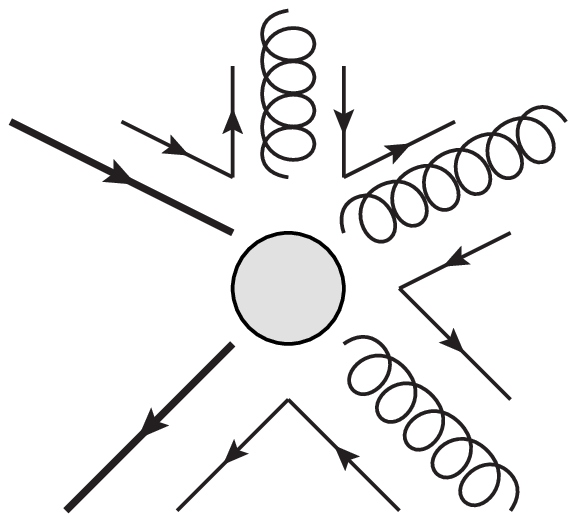}
\end{tabular}

Gluon emission off a non-circular chain.\\ The chain stays non-circular.

\begin{tabular}{m{1.5cm}m{0.5cm}m{1.5cm}m{0.5cm}m{1.5cm}}
\includegraphics[scale=0.3]{circular_chain}\hspace*{1cm} & {\huge $\to$} &
\includegraphics[scale=0.3]{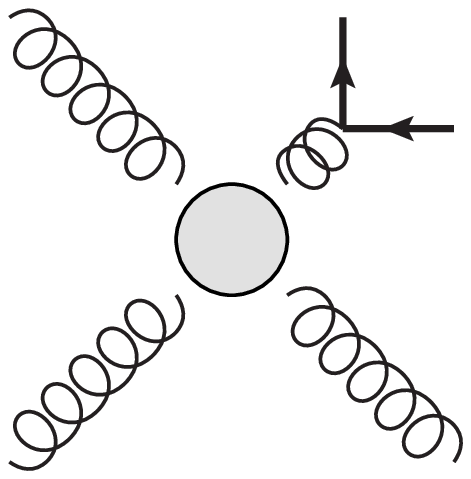}\hspace*{1cm} & {\huge $\to$} &
\includegraphics[scale=0.3]{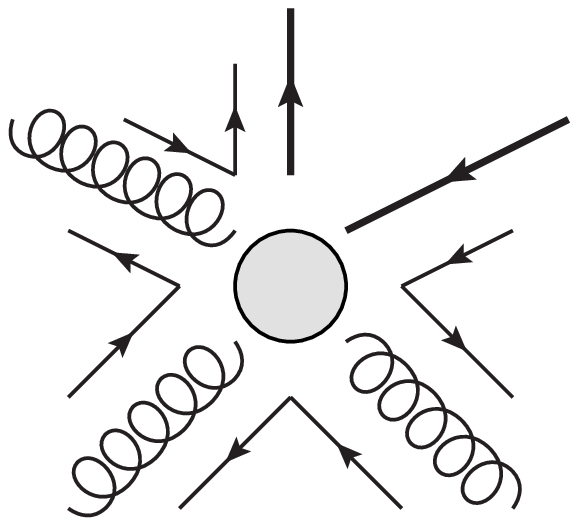}
\end{tabular}

$g\to q\bar{q}$ splitting in a circular chain.\\ The chain becomes non-circular.

\begin{tabular}{m{1.5cm}m{0.5cm}m{1.5cm}m{0.5cm}m{1.5cm}}
\includegraphics[scale=0.3]{non_circular_chain}\hspace*{1cm} & {\huge $\to$} &
\includegraphics[scale=0.3]{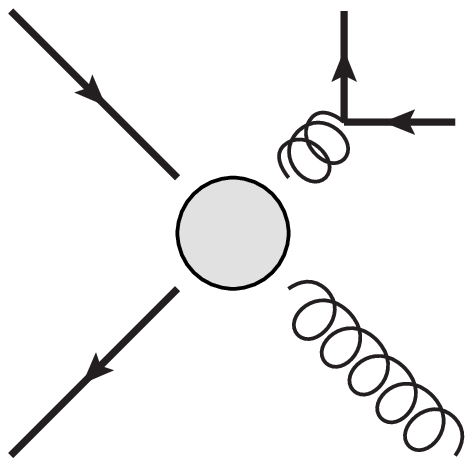}\hspace*{1cm} & {\huge $\to$} &
\includegraphics[scale=0.3]{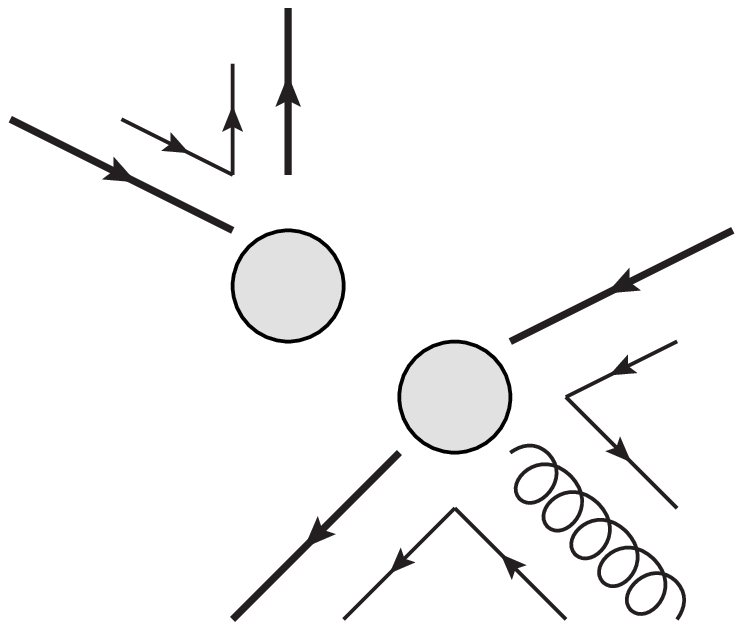}
\end{tabular}

$g\to q\bar{q}$ splitting in a non-circular chain,\\ triggering
breakup of the chain.

\end{center}

\caption{\label{figures:chainsplittings}Examples of parton emission
  from dipole chains. In these examples always the upper dipole has
  been considered for emissions. Note that any dipole may split in two
  different ways, splitting either of its legs. These competing
  possibilities are not shown in the transition diagrams.}
\end{figure}

As the evolution factors into dipole chains as independently evolving
objects, all possible emitters in the chain -- after having inserted
the generated splitting -- now get the selected $q_\perp^2$ assigned
as their hard scale, or stay at the kinematically allowed scale
$p_{\perp,i,j}^2$ if $q_\perp^2 > p_{\perp,i,j}^2$.  If a $g\to
q\bar{q}$ splitting has been selected for a circular chain, this chain
becomes non-circular. If it has been selected for an already
non-circular chain, this chain breaks up into two independent chains
exactly between the $q\bar{q}$-pair, owing to the colour structure of
this splitting. This situation, along with non-exceptional splittings
is depicted in Fig.~\ref{figures:chainsplittings}.

\subsection{Finishing the Shower}

After the shower evolution has terminated, the incoming partons with
momenta $p_{a,b}$ in general have non-vanishing transverse momenta
with respect to the beam directions. This necessitates a realignment
of the complete event encountered at this stage.  Following the
arguments of \cite{Platzer:2009jq}, the momenta of the evolved
incoming partons $p_{a,b}$ are taken to {\it define} the frame of the
collision at hand, {\it i.e.} hadron momenta $\tilde{P}_{a,b}$.  We
then seek a Lorentz transformation to take $\tilde{P}_{a,b}$ to the
externally fixed hadron momenta $P_{a,b}$, which is in turn used to
realign the complete event.

To construct the momenta of the incoming hadrons $\tilde{P}_{a,b}$, we
require the three-momenta of $\tilde{P}_{a,b}$ being collinear with the
respective partonic three-momenta and define momentum fractions
\begin{equation}
x_{a,b} = \frac{2 \tilde{P}_{b,a}\cdot p_{a,b}}{S} \ .
\end{equation}
The momentum fractions are further constrained by requiring that
\begin{equation}
(\tilde{P}_a+\tilde{P}_b)^2 = S
\end{equation}
where $S$ is the centre-of-mass energy squared of the collision, such
that the desired Lorentz transformation exists.

The second constraint is in principle to be chosen in such a way as to
preserve the most relevant kinematic quantity of the hard process
which initiated the showering. By default, we choose this to be the
rapidity of a system $X$, which is either the system of non-coloured
particles at the hard sub-process, or the complete final state in case
of a pure QCD hard scattering.

\subsection{Cluster Hadronization}

The cluster hadronization model, originally proposed in
\cite{Webber:1983if}, is the hadronization model used by the
\hpp{} event generator.  The model in its initial stage
just after parton showering, performs a splitting of gluons into
quark-antiquark pairs such that in the large-$N_c$ limit a set of
colour singlet clusters emerge from the event under consideration.

These clusters are then subsequently converted into hadrons, by either
splitting them into clusters of lower invariant mass or performing
directly the decay to meson pairs, in case another $q\bar{q}$ pair is
`popped' from the vacuum inside the cluster, or baryon pairs, where
the creation of a diquark-antidiquark pair is assumed. Further details
of the model will not be discussed here.

The main assumption of the model is however, that both quarks are
located on their {\it constituent} mass shell, and gluons are as well
assigned a non-vanishing constituent mass, entering as a parameter of
the model.  In the standard \hpp{} parton shower, acting as a $1\to
2$ cascade, only scales and momentum fractions of the splittings are
determined during the evolution, the full kinematic information being
constructed after the end of the perturbative evolution. This setup
thus straightforwardly allows to include the constituent masses in
this particular step.  Since the dipole shower preserves momentum
conservation locally to each splitting, ending up with a set of
massless partons, such a treatment is not possible.

The way to perform the `reshuffling' of the massless parton momenta to
their constituent mass shells is chosen to be the following algorithm:
Let $Q_c$ be the total momentum of all final state partons and perform
a boost $\Lambda_c$ to the centre-of-mass system of $Q_c$, $\Lambda_c
Q_c = (\hat{Q}_c,{\mathbf 0})$.  The boosted parton momenta $p_i$ are
now put on the constituent mass shell, including a global rescaling of
their three-momenta,
\begin{equation}
p_i = \left(|{\mathbf p}_i|,{\mathbf p}_i\right)\to 
p'_i = \left(\sqrt{\xi^2 |{\mathbf p}_i|^2+m_{c,i}^2},\xi {\mathbf p}_i\right) \ .
\end{equation}
Momentum conservation then implies the following relation be satisfied,
\begin{equation}
\hat{Q}_c = \sum_i \sqrt{\xi^2 |{\mathbf p}_i|^2+m_{c,i}^2} \ ,
\end{equation}
which may be solved numerically to yield a value for $\xi$. Finally
the inverse boost $\Lambda_c^{-1}$ is applied to the new parton
momenta $p'_i$.

\section{The Matchbox Implementation}
\label{sections:matchbox}

\begin{figure}
\centering
\includegraphics[scale=0.5]{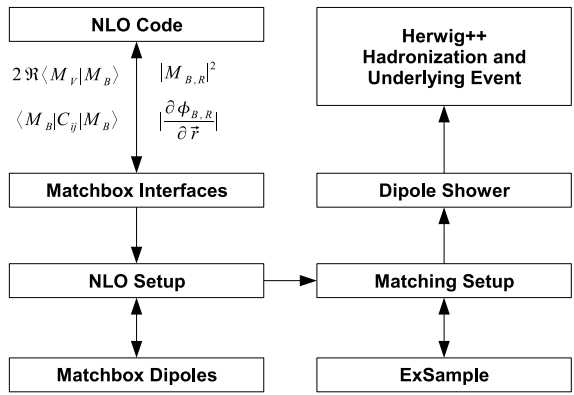}
\caption{\label{figures:flow-diagram}A sketch of the interaction of
  the Matchbox and dipole shower modules as integrated in
  \hpp{}. To perform a matched NLO calculation an external
  code only has to provide tree-level and one-loop amplitudes along
  with colour- and spin-correlated amplitudes of the Born process and
  an appropriate phase space generator.}
\end{figure}

Closely related to the dipole shower implementation, though
technically independent of it, is the development of the
\matchbox{} module. \matchbox{} is based on an extended
version of \thepeg{}, the extensions providing functionality to
perform hard process generation at the level of NLO QCD accuracy and
easing the setup of run time interfaces to external codes for hard
process generation.  We have implemented an automated generation of
subtraction terms based on the dipole subtraction formalism
\cite{Catani:1996vz}, based on the information available from
\thepeg{} matrix element implementations, which will be
discussed in further detail in section~\ref{sections:dipoles}. A full
NLO calculation to be run in the \matchbox{} framework only
requires the implementation of tree-level and one-loop amplitudes, the
presence of colour (and spin) correlated amplitudes for the Born
process and the presence of a phase space generator appropriate to the
process under consideration. Fig.~\ref{figures:flow-diagram} sketches
the involved software modules and their interaction with an external
implementation of a NLO calculation.

Besides being capable of performing a Monte Carlo integration of
`plain' NLO corrections, the main purpose of \matchbox{} is to
turn a NLO calculation into a matched calculation to be consistently
combined with a parton shower. Here, functionality is especially
provided to calculate the inclusive NLO cross section differential in
the Born degrees of freedom, which, along with a matrix element
correction to the shower, is the main ingredient to the \powheg{} method
of combining parton showers and NLO QCD corrections.

\matchbox{} is automatically generating matrix element
corrections from the NLO real emission contribution.  It further
allows the possibility to overcome problems in the \powheg{} matching
owing to radiation zeroes in the Born matrix element.  The matrix
element correction splitting kernel, which is essentially defined by
the ratio of real emission and Born matrix elements squared is turned
into the corresponding distribution including the Sudakov form factor
by using the \exsample{} library, \cite{Platzer:2011dr}.
\exsample{} allows the efficient sampling of distributions of
this type, without having to provide any analytic knowledge on the
splitting kernel or trying to estimate enhancement factors to simpler
functions such as dipole splitting kernels. \exsample{} is
also used to sample emissions in the dipole shower implementation.

\subsection{Notation}
\label{sections:notation}

We consider NLO calculations carried out using the dipole subtraction
method, \cite{Catani:1996vz}.  Instead of using the notation
established there, we unify the indices of all possible dipoles to
ease readability, as expressions become quite complicated especially
when considering the \powheg{} type matching. For the subtraction dipoles
we choose the notation
\begin{equation}
{\cal D}_{ij,k}\ , {\cal D}_{ij}^a\ , {\cal D}_k^{ai}\ ,
{\cal D}^{ai,b}\quad \to\quad {\cal D}_\alpha \ ,
\end{equation}
where the arguments are unified and we make explicit the dependence
on either real emission or `tilde' kinematics, {\it e.g.}
\begin{equation}
{\cal D}_{ij,k}(q_a,q_b;q_1,...,q_{n+1}) \quad\to\quad {\cal D}_\alpha(p_n^\alpha(q_{n+1})|q_{n+1}) \ .
\end{equation}
In this notation, $p_{n}$ now refers to the whole phase space point,
\begin{equation}
p_a,p_b;p_1,...,p_n \quad\to\quad p_n\equiv (\hat{p}_a,\hat{p}_b;\hat{p}_1,...,\hat{p}_n) \ ,
\end{equation}
where we have added hat symbols to the momenta to distinguish a single
momentum from a complete phase space point.  The `tilde' mapping and
its inverse are denoted by
\begin{align}
\widetilde{p_{ij}}(q_i,q_j,q_k)\ , \widetilde{p_k}(q_i,q_j,q_k)
&\to p_n^\alpha(q_{n+1}) \\\nonumber
q_{i,j,k}(\widetilde{p_{ij}},\tilde{p_k};p_\perp^2,z,\phi)
&\to q_{n+1}^\alpha(p_n;p_\perp^2,z,\phi) \ .
\end{align}
Differential cross sections are considered in collinear factorisation,
\begin{multline}
{\rm d}\sigma_X(p_n|Q,x_a,x_b,\mu_F) = \\f_{P\gets a}(x_a,\mu_F)f_{P\gets b}(x_b,\mu_F)
{\rm d}\sigma_X(p_n|Q){\rm d}x_a {\rm d}x_b
\end{multline}
where the partonic cross section is in general of the form
\begin{equation}
{\rm d}\sigma_X(p_n|Q) = F(\hat{p}_a,\hat{p}_b) X(p_n) {\rm d}\phi(p_n|Q) \ .
\end{equation}
Here $F(\hat{p}_a,\hat{p}_b)$ is the appropriate flux factor and
$X(p_n)$ generically denotes any contribution to the cross section
which can be cast in the above form, {\it i.e.} tree-level amplitudes
squared, one-loop tree-level interferences, subtraction terms, or the
`deconvoluted' finite collinear terms to be discussed below.  The
phase space measure ${\rm d}\phi(p_n|Q)$ is given by
\begin{multline}
{\rm d}\phi(p_n|Q) = \\
(2\pi)^d\delta\left(\sum_{i=1}^n p_i-p_a-p_b-Q\right)
\prod_{i=1}^n \frac{{\rm d}^{d-1}\hat{\mathbf{q}}_i}{(2\pi)^{d-1} 2 \hat{q}_i^0}
\end{multline}
In latter sections, it will turn out to be useful to rewrite
this as
\begin{align}\nonumber
{\rm d}\sigma_X(p_n|Q,x_a,x_b) & = X(p_n){\rm d}F(x_a,\hat{p}_a,x_b,\hat{p}_b){\rm d}\phi(p_n|Q)
\\
&\equiv X(p_n) {\rm d}\phi_F(p_n|Q,x_a,x_b) \ ,
\end{align}
where we dropped making explicit the factorisation scale dependence
from now on.

The finite collinear terms originating from counter terms to
renormalise parton distribution functions and integrated subtraction
terms are reported in \cite{Catani:1996vz}. These are given as
convolutions of Born-type cross sections of colour correlated
amplitudes with certain `insertion operators', {\it e.g.} for the
incoming parton $a$
\begin{equation}
\int_0^1 {\rm d}z\ C(p_n^a(z)){\rm d}\phi(p_n|Q^a(z)){\rm d}F(x_a,z\hat{p}_a,x_b,\hat{p}_b) \ ,
\end{equation}
where the superscript $a$ along with an argument $z$ indicates, that
parton $a$'s momentum is rescaled by $z$. The insertion operators
themselves include $+$-distributions, and events should be generated
according to the rescaled incoming momentum $z\hat{p}_a$. A numerical
implementation is at first sight not obvious.  Considering however the
integration over the momentum fraction $x_a$, these contributions can
be rewritten in terms of a Born-type cross section multiplied by
modified PDFs along the lines of
\begin{multline}
\int_0^1{\rm d}x \int_0^1 {\rm d}z f(x)B(x z)P(z) = \\
\int_0^1{\rm d}x B(x) \int_x^1 \frac{{\rm d}z}{z} f\left(\frac{x}{z}\right)P(z)
\end{multline}
and the $+$-distributions can be expressed in a way to allow for
numerical implementation.  All possible contributions for light quarks
are implemented in \matchbox{}.

Any NLO cross section within the dipole subtraction thus takes the form
\begin{align}
\label{equations:dipolenlo}
\sigma_{NLO}
& = \int |{\cal M}_B(p_n)|^2 u(p_n) {\rm d}\phi_F(p_n|Q,x_a,x_b)
\\\nonumber
& +\int \left[2 {\rm Re}\langle {\cal M}_B^*(p_n) {\cal M}_V(p_n)\rangle + \right.\\\nonumber
&\qquad\left.
\langle {\cal M}_B(p_n)|{\mathbf I}|{\cal M}(p_n)\rangle\right]_{\epsilon=0} u(p_n) {\rm d}\phi_F(p_n|Q,x_a,x_b)
\\\nonumber
& + \int
\langle {\cal M}_B(p_n)|(\tilde{{\mathbf P}}+\tilde{{\mathbf K}})|{\cal M}(p_n)\rangle u(p_n) 
{\rm d}\tilde{\phi}_F(p_n|Q,x_a,x_b)
\\\nonumber
& + \int \left(|{\cal M}_R(q_{n+1})|^2 u(q_{n+1})\right.\\ \nonumber
&\qquad - \left.\sum_\alpha
{\cal D}_\alpha(p_n^\alpha(q_{n+1})|q_{n+1}) u(p_n^\alpha(q_{n+1})) \right)\\\nonumber
& \qquad\times{\rm d}\phi_F(q_{n+1}|Q,x_a,x_b)
\end{align}
where the insertion operators ${\mathbf I}$ are given in
\cite{Catani:1996vz} and have been implemented for light quarks in
full generality as well.  $\tilde{{\mathbf P}}$, $\tilde{{\mathbf K}}$
and ${\rm d}\tilde{\phi}_F$ denote the deconvoluted versions of the
finite collinear terms originating from the insertion operators
${\mathbf P}$,${\mathbf K}$ given in \cite{Catani:1996vz}.  Here, the
test functions $u(p_n)$ refer to the class of events to be generated
by a Monte Carlo realisation of the above integrals, and ${\cal
  M}_{B,R}$ denote the Born and real emission amplitudes,
respectively.  Since only the structure of the real emission and
subtraction terms turns out to be relevant for matching purposes, we
from now on collectively denote Born, virtual and insertion operator
contributions by
$$
\int |{\cal M}_{BV}(p_n)|^2 u(p_n) {\rm d}\phi_F(p_n|Q,x_a,x_b) \ .
$$

Since all the integrals will be dealt with by means of Monte Carlo
methods, differentials are expressed in terms of a Jacobian expressing
the physical variables in terms of random numbers and a volume element
on the unit hypercube of these random numbers, {\it e.g.}
\begin{equation}
{\rm d}\phi(p_n|Q) = \left|\frac{\partial p_n}{\partial \vec{r}} \right|{\rm d}^k r \ .
\end{equation}
We identify ratios of differentials to actually mean the ratios
of the corresponding functions multiplied by the Jacobian in use
to express them in terms of random numbers, {\it e.g.} for two cross sections
we define
\begin{equation}
\frac{{\rm d}\sigma_X(q_m|Q)}{{\rm d}\sigma_Y(p_n|Q)} \equiv
\frac{X(q_m)}{Y(p_n)} 
\frac{\left|\frac{\partial q_m}{\partial \vec{r}_q} \right|}
{\left|\frac{\partial p_n}{\partial \vec{r}_p} \right|} \ .
\end{equation}

\subsection{Automated Dipole Subtraction}
\label{sections:dipoles}

Any matrix element implemented in \thepeg{} is expected to
provide information on the diagrams contributing to it.  It is this
information, which is used to generate subtraction dipoles by a simple
algorithm of checking, for any contributing diagram, if any two
external coloured legs are attached to the same vertex. By removing
this vertex from the diagram information, the diagram of the
corresponding `underlying Born process' is obtained.  Conversely, the
same pairing of diagrams provides a way to identify which real
emission processes are to be considered given any Born process. This
information is used when setting up the inclusive NLO cross section
calculation and generating matrix element corrections for the parton
shower.  From a given matrix element object implementing a real
emission contribution, \matchbox{} checks a set of Born matrix
element objects provided along with the real emission ones for the
underlying Born processes obtained and adds all matching pairs to the
calculation if there exists a subtraction dipole object which claims
responsibility for the given pairing.  Similarly, all insertion
operator implementations present are checked if they claim
responsibility for a given Born process, thus completing the setup of
a NLO calculation.  The complete calculation is then injected as a
\thepeg{} \program{SubProcessHandler} object into the stage of
event generation.

For running unmatched calculations, a group of events consisting of
real emission and `tilde' phase space points is provided along with
the relative weights of the individual contributions present in the
group. The sum of these weights, {\it i.e.} real emission minus
subtraction term contributions is driving the cross section
integration and event unweighting.

\subsection{Subtractive NLO Matching}
\label{sections:subtractive}

Owing to the fact that the dipole shower implementation uses splitting
kernels which precisely equal the dipole subtraction terms, following
the steps leading to MC@NLO here results in a very simple
matching.\footnote{Though the kinematic parametrisation differs from
  the one used in the subtraction context, it can be related to the
  usual `tilde' parametrisation by a boost in case a single emission
  is considered.} This subtractive matching is basically identical to
the NLO calculation itself, except that instead of event groups now a
single real emission phase space point is generated from the
subtracted real emission contribution.  In an algorithmic manner, the
matching may thus be expressed very simply:
\begin{itemize}
\item Generate Born-type events $p_n$ with density
\begin{equation}
|{\cal M}_{BV}(p_n)|^2 {\rm d}\phi_F(p_n|Q,x_a,x_b) \ ,
\end{equation}
\item generate real-emission type events $q_{n+1}$ with
density
\begin{multline}
\left(|{\cal M}_R(q_{n+1})|^2 
- \sum_\alpha {\cal D}_\alpha(p_n^\alpha(q_{n+1})|q_{n+1}) \right)\\\times
{\rm d}\phi_F(q_{n+1}|Q,x_a,x_b) \ ,
\end{multline}
\item and feed either into the dipole shower.
\end{itemize}
A subtlety, however, arises here. Since we are interested in
describing the hardest emission according to the exact real emission
matrix element, the parton shower should not generate harder emissions
than the one fixed from the NLO calculation. Practically, this is
implemented by calculating the $p_\perp^\alpha$ as defined by the
inverse `tilde' mapping from each dipole configuration $\alpha$, since
the kinematics of the emission appears differently depending on the
emitting dipole considered. $p_\perp^\alpha$ is communicated as a veto
scale to the dipole shower, which is not allowed to generate emissions
with $p_\perp > p_\perp^\alpha$ off the emitter, emission and
spectator partons used to evaluate ${\cal D}^\alpha$. Another
approach, in which the dipole shower is generally not allowed to emit
at scales $p_\perp$ larger than final state transverse momenta can
equivalently be used and may become the default in a future
version. This treatment is then very similar to the \herwig{} shower
in use with the traditional MC@NLO implementation.

\subsection{NLO Matching with Matrix Element Corrections}
\label{sections:powheg}

The splitting kernels to be used for a matrix element correction are
given by the ratio of real emission and Born matrix elements squared,
weighted by (in principle) arbitrary weight functions for each
kinematic mapping of a subtraction term, {\it i.e.} for each
subtraction term. It is most simple to choose the subtraction terms
themselves to define these weight functions. This has the advantage
that all divergences but the divergence associated to the subtraction
term ${\cal D}_\alpha$ are divided out from the real emission matrix
element, and dynamical features of the Born matrix element, like peaks
owing to unstable particles, are flattened out in the splitting kernel
considered.

Within this procedure, one faces three major problems:
\begin{itemize}
\item Some of the subtraction dipoles, in particular the ones with
  initial state emitter and final state spectator or vice versa, are
  not positive-definite. This makes a Monte Carlo treatment of the
  corresponding Sudakov-type distribution hard to implement.  Since
  the regions, where these dipole kernels become negative correspond
  to hard, large angle parton emission, it is clear that this problem
  can be cured by changing the irrelevant finite terms of the
  subtraction dipoles, provided they are consistently taken into
  account in the integrated ones. Within the \matchbox{}
  implementation this has so far been carried out for the $qq$
  initial-final dipoles, which have been modified to reproduce the
  matrix element squared for gluon emission off the corresponding
  vector current and are thus positive by definition.
\item The Born matrix element squared may contain zeroes.
  In this case, its inverse is obviously ill-defined.
\item The implementation of the parton densities at hand, which enter
  as a ratio in the splitting kernels as well, may not be stable in
  particular for large $x$ in the sense that the interpolation used
  oscillates around zero rather than tending to zero smoothly. This
  poses a problem similar to the zeroes in the Born matrix element,
  however now without any physical interpretation.
\end{itemize}

The latter two problems can be solved by introducing an auxiliary
cross section ${\rm d}\sigma_{\text{screen}}(p_n|Q;p_\perp^2)$ which
enters into the definition of the splitting kernels
\begin{multline}
\label{equations:powhegkernel}
{\rm d}P_\alpha(p_\perp^2,z,\phi|p_n) =
{\rm d}^3r \frac{{\cal D}_{\alpha}(p_n|q_{n+1}^\alpha)}
{\sum_\beta {\cal D}_{\beta}(p_n^\beta(q^\alpha_{n+1})|q_{n+1}^\alpha)}\\\times
\frac{{\rm d}\sigma_R(q_{n+1}^\alpha|Q,x_a',x_b')}
{{\rm d}\sigma_B(p_n|Q,x_a,x_b)+{\rm d}\sigma_{\text{screen},\alpha}(p_n|Q;p_\perp^2)} \ ,
\end{multline}
where we have already written the splitting kernel differential in the
random numbers determining $p_\perp^2$, $z$ and $\phi$, and the
dependence of $q^\alpha_{n+1}=q^\alpha_{n+1}(p_n;p_\perp^2,z,\phi)$ on
the splitting variables is understood implicitly. In order not to change
the divergence structure implying the resummation of large logarithms,
the screening cross section needs to vanish as $p_\perp^2\to 0$.
Since Born zeroes cannot occur for $p_\perp^2\to 0$ (the QCD
singularities factor in this limit with respect to the Born process)
Eq.~\eqref{equations:powhegkernel} is free of these problems. If, in
addition, the screening cross section does not depend on the parton
distributions, the technical issues with PDFs becoming zero are cured
as well.

The screening cross section has however to be taken into account for
the fixed order calculation in order to reproduce the correct NLO
cross section and will thereby spoil the original simplicity of using
the NLO $K$-factor differential in the Born variables to generate
events to enter the matrix element corrected shower.  Including the
screening cross section the fixed order cross section can then be
calculated to be constructed of densities for Born-type and real
emission type events. The densities for Born-type events closely
resemble the $K$-factor modification,
\begin{multline}
{\rm d}\sigma_{\text{inclusive}}(p_n|Q,x_a,x_b)
= \\{\rm d}\sigma_{BV}(p_n|Q,x_a,x_b) +
\int{\rm d}^3 r \frac{{\rm d}\sigma_{\text{R,inclusive}}(p_n|Q,x_a,x_b)}{{\rm d}^3 r}
\end{multline}
where
\begin{multline}
\frac{{\rm d}\sigma_{\text{R,inclusive}}(p_n|Q,x_a,x_b)}{{\rm d}^k r_B {\rm d}^3 r} = \\
\frac{{\rm d}\sigma_B(p_n|Q)}{{\rm d}^k r_B}
\sum_\alpha \frac{{\cal D}_{\alpha}(p_n|q_{n+1}^\alpha)}
{\sum_\beta {\cal D}_{\beta}(p_n^\beta(q^\alpha_{n+1})|q_{n+1}^\alpha)} 
R(p_n|q_{n+1}^\alpha) \ ,
\end{multline}
and
\begin{multline}
R(p_n|q_{n+1}^\alpha) =
- \frac{{\rm d}\phi_F(q_{n+1}^\alpha|Q,x_a',x_b')}{{\rm d}\phi(p_n|Q)}\\ +
\frac{{\rm d}\sigma_R(q_{n+1}^\alpha|Q,x_a',x_b')}
{{\rm d}\sigma_B(p_n|Q,x_a,x_b)+{\rm d}\sigma_{\text{screen},\alpha}(p_n|Q;p_\perp^2)} \ .
\end{multline}
To generate events according to these densities, a $k+3$-dimensional
random number point is chosen, where the three additional degrees of
freedom are discarded. Owing to the fact that the integration volume
in terms of random numbers is the unit hypercube, this procedure
produces the integration over the degrees of freedom of the parton
emitted in the real emission on average.

Events of real emission type are to be generated with density
\begin{multline}
{\rm d}\sigma_R(q_{n+1}|Q,x_a,x_b)\ \times \\
\sum_\alpha \bar{R}(p_n^\alpha|q_{n+1})
\frac{{\cal D}_{\alpha}(p_n^\alpha|q_{n+1})}
{\sum_\beta {\cal D}_{\beta}(p_n^\beta|q_{n+1})} \ ,
\end{multline}
\begin{multline}
\bar{R}(p_n^\alpha|q_{n+1}) = \\
\frac{{\rm d}\sigma_{\text{screen},\alpha}(p_n^\alpha|Q;p_\perp^2)}
{{\rm d}\sigma_B(p_n^\alpha|Q,x'_a,x'_b)+{\rm d}\sigma_{\text{screen},\alpha}(p_n^\alpha|Q;p_\perp^2)} \ ,
\end{multline}
which is just a reweighting of the real emission contribution.  Events
of both classes can then be showered by a parton shower using a matrix
element correction as defined at the beginning of this section, and a
communication of veto scales applies to the real emission contribution
along the same lines as for the subtractive matching. Note that the
individual contributions are positive, as long as the screening cross
section is bounded from above by a reasonable value.

Since this type of matching is independent of the parton shower to act
downstream, the actual implementation does not make any reference to
the dipole parton shower, and real emission contributions according to
the matrix element correction are generated outside any shower module,
presenting a real emission sub process supplemented with proper veto
scales, or a Born-type sub process to the shower, if radiation has
been generated according to the matrix element correction or not,
respectively.

Note that, when putting the screening cross section to zero, the
original simplicity of the \powheg{}-type matching is recovered. The
matrix element corrections, inclusive and real-emission type
contributions are all setup and calculated in an automated way within
the \matchbox{} implementation. The screening cross section is
by default chosen from the corresponding phase space and the
dimensionality required by the phase space, {\it i.e.}
\begin{equation}
{\rm d}\sigma_{\text{screen},\alpha}(p_n^\alpha(q_{n+1})|Q;p_\perp^2) =
\frac{(p_\perp^\alpha)^2}{s_{\alpha}(q_{n+1})} \frac{{\rm d}\phi(q_{n+1}|Q)}{(s_{\alpha}(q_{n+1}))^{n_{\text{out}}}} \ ,
\end{equation}
where $p_\perp^\alpha$ is the transverse momentum associated to the
mapping $p_n^\alpha(q_{n+1})$, $s_{\alpha}(q_{n+1})$ is the
appropriate mass squared of the emitter-spectator pair in
$p_n^\alpha$, and $n_{\text{out}}$ is the number of outgoing
particles. Other choices may be possible.

\section{Results at LEP}
\label{sections:lep}

The variety of data acquired by the LEP experiments allow for a
systematic fit of parameters of the parton shower and the
hadronization model.  In a preliminary fit, the parameters assumed to
mainly determine the description of event shape variables and jet
rates as measured by the DELPHI experiment \cite{Abreu:1996na} and jet
observables as reported by the OPAL collaboration
\cite{Pfeifenschneider:1999rz} have been fitted using the
\rivet{} \cite{Buckley:2010arxb} and \professor{}
\cite{Buckley:2009bj} systems. The parameters and ranges considered
are given in Tab.~\ref{tables:lepfitparameters}, along with a short
description. Parameters which are known to mainly affect individual
hadron multiplicities have not been varied, and fragmentation
parameters for heavy quarks have been set equal to the values of those
for light quarks. A simple modification of the running of $\alpha_s$
in the infrared has been adopted by replacing its argument $q^2\to
q^2+\mu_{\text{soft}}^2$. This modification has originally been
motivated to supply another model for intrinsic transverse momentum
generation by letting the initial state shower evolve down to very
small scales along the lines of \cite{Gieseke:2007ad}. We see however
no reason that it should not be considered for final state radiation
as well.

Separate fits have been performed for LO and NLO predictions. LO
predictions have been obtained by running just the parton shower,
using a one-loop running $\alpha_s$. NLO prediction have been obtained
by means of supplementing the shower with the matrix element
correction matching without using the Born screening cross section and
a two-loop running $\alpha_s$. In total we find that the NLO
simulation gives a marginally better fit than the LO one, though the
description of data is completely comparable within experimental
uncertainties.

\begin{figure}
\centering
\includegraphics[scale=0.6]{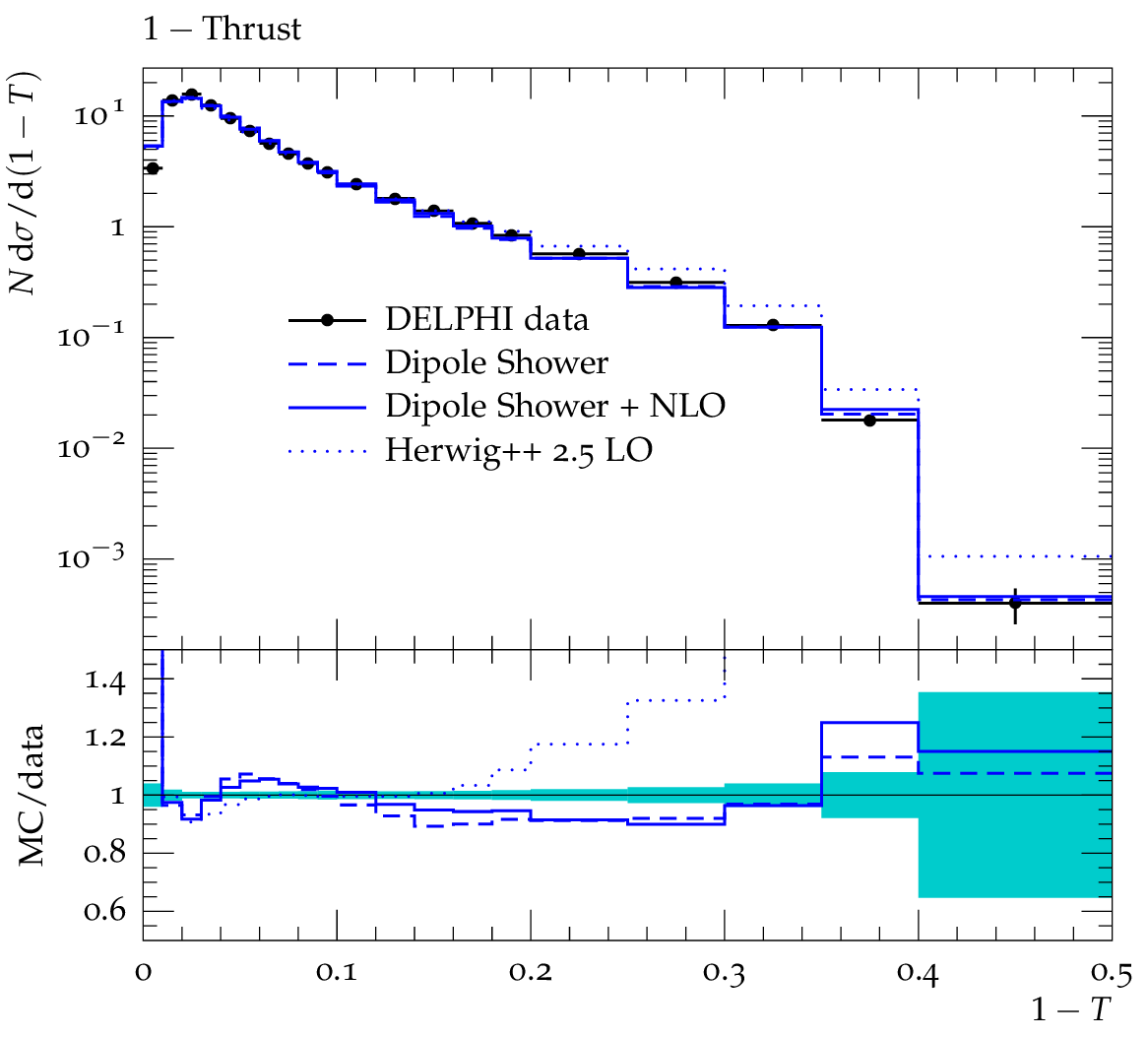}
\includegraphics[scale=0.6]{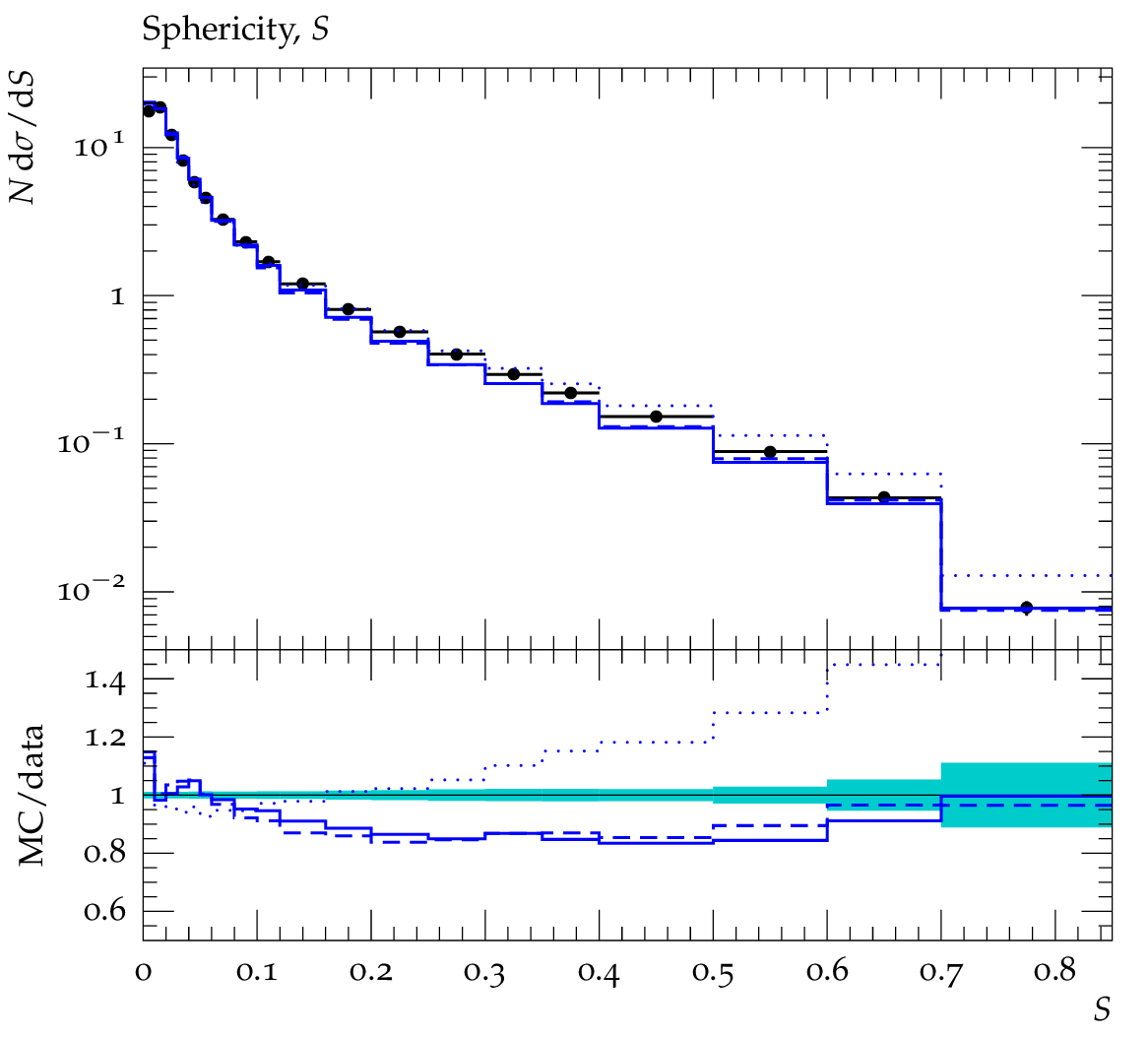}
\includegraphics[scale=0.6]{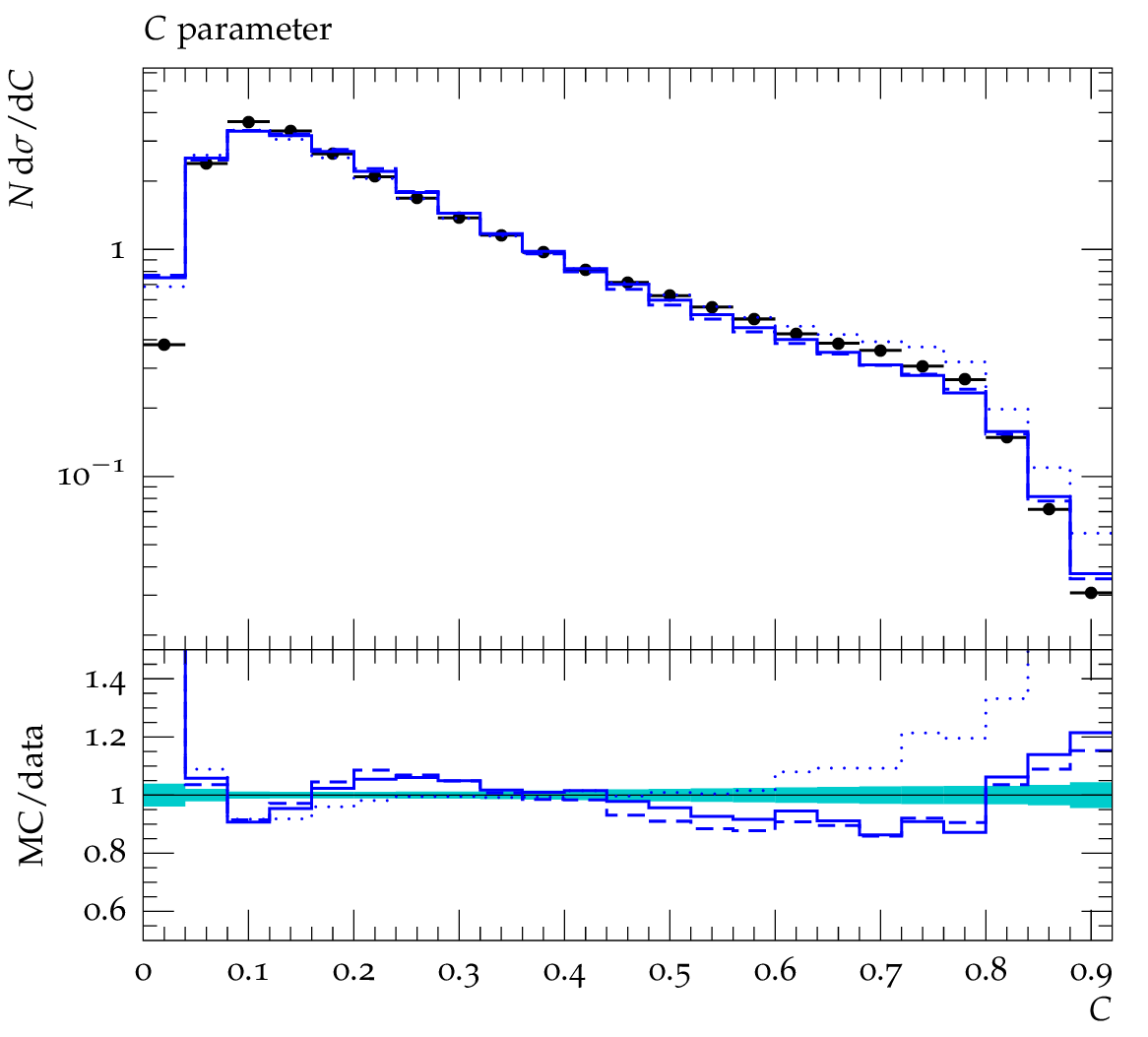}
\caption{\label{figures:lepshapes-lovsnlo}Some event shape variables
  as predicted by the leading order and next-to-leading order
  simulations. Here, we additionally compare to the standard \hpp{}
  shower (version 2.5.1 with default settings), showing that the
  dipole shower gives a significantly improved description already at
  leading order.}
\end{figure}

\begin{figure}
\centering
\includegraphics[scale=0.6]{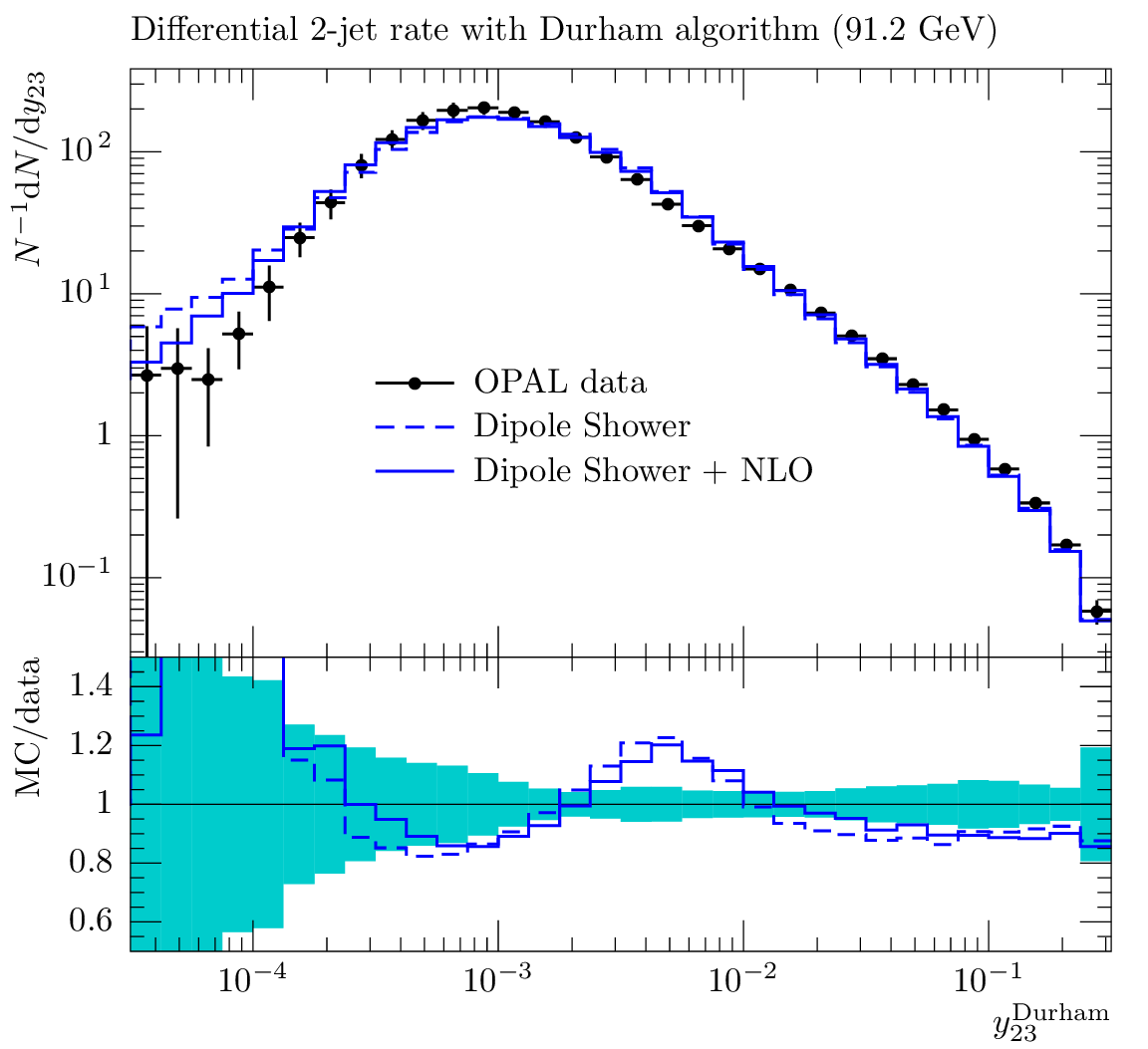}
\caption{\label{figures:lepjets-lovsnlo}The differential three jet
  rate as predicted by the leading order and next-to-leading order
  simulations.}
\end{figure}

\begin{figure}
\centering
\includegraphics[scale=0.6]{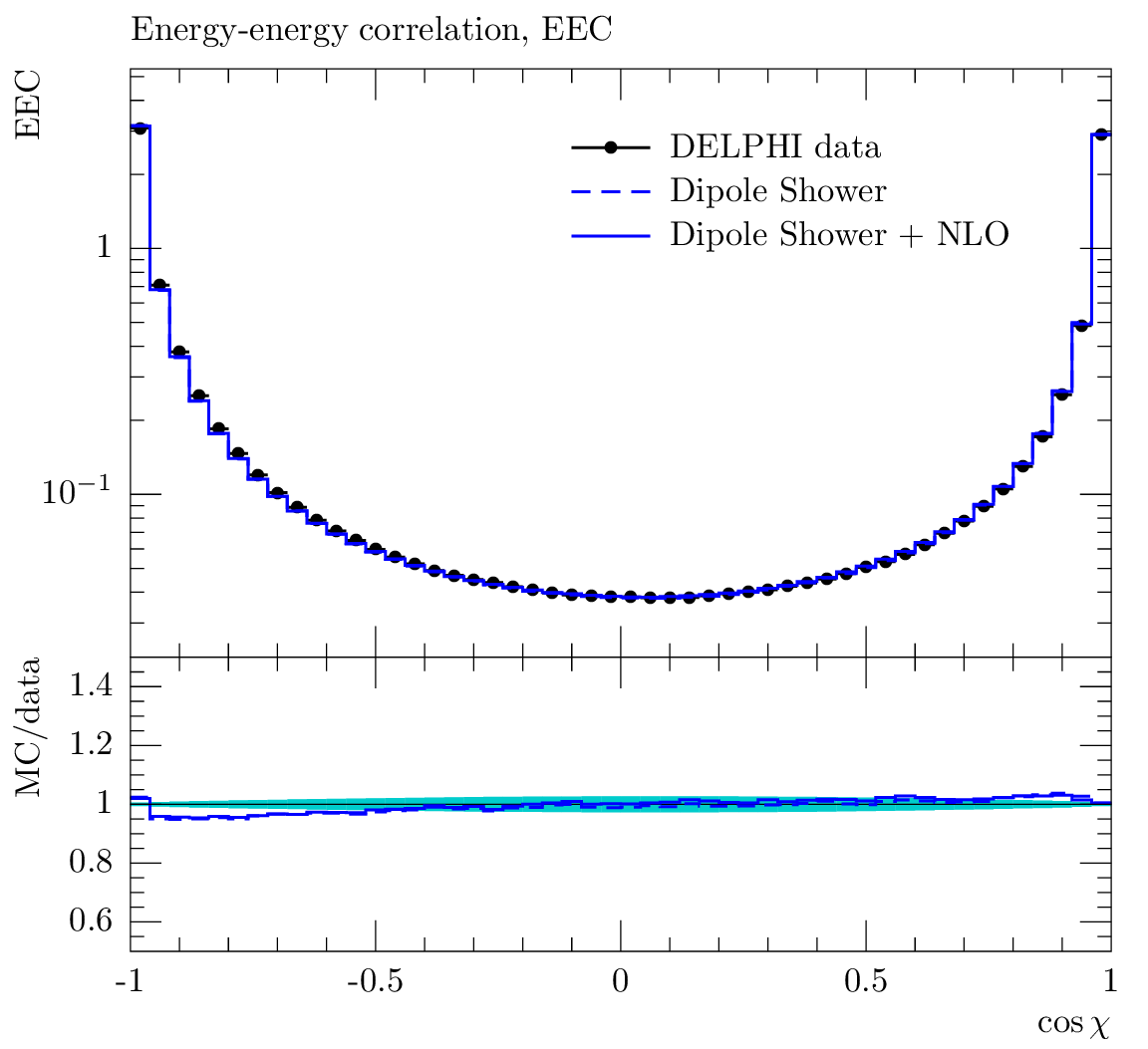}
\caption{\label{figures:lepeec-lovsnlo}Energy-energy correlation.
  Note that this observable has not been included in the fit.}
\end{figure}

The fitted parameter values are displayed in
Tab.~\ref{tables:lepfittedparameters}. Most notably, the
hadronization parameters for the LO and NLO fit do not significantly
differ.  For both predictions, a modification of the infrared running
of $\alpha_s$ seems not to be preferred.  The infrared cutoff of the
parton shower is determined more precisely by the NLO fit, which
prefers a smaller cutoff. Also $\alpha_s(M_Z^2)$ is determined more
precisely by the NLO fit. Both $\alpha_s$ values obtained are
compatible with the world average \cite{PDG:2010} of $0.1184$, where
the NLO result is closer to this value. Note that this should be
regarded a coincidence at the level of the approximation considered
and it is certainly not possible to uniquely relate the obtained
value to one applying to the $\overline{MS}$ scheme. In
Figs.~\ref{figures:lepshapes-lovsnlo} and
\ref{figures:lepjets-lovsnlo} the LO and NLO simulation results are
compared for selected observables. Fig.~\ref{figures:lepeec-lovsnlo}
shows the energy-energy-correlation, which has not been included in
the fit.

\subsection{Comparison of Matching Strategies}

\begin{figure}
\centering
\includegraphics[scale=0.6]{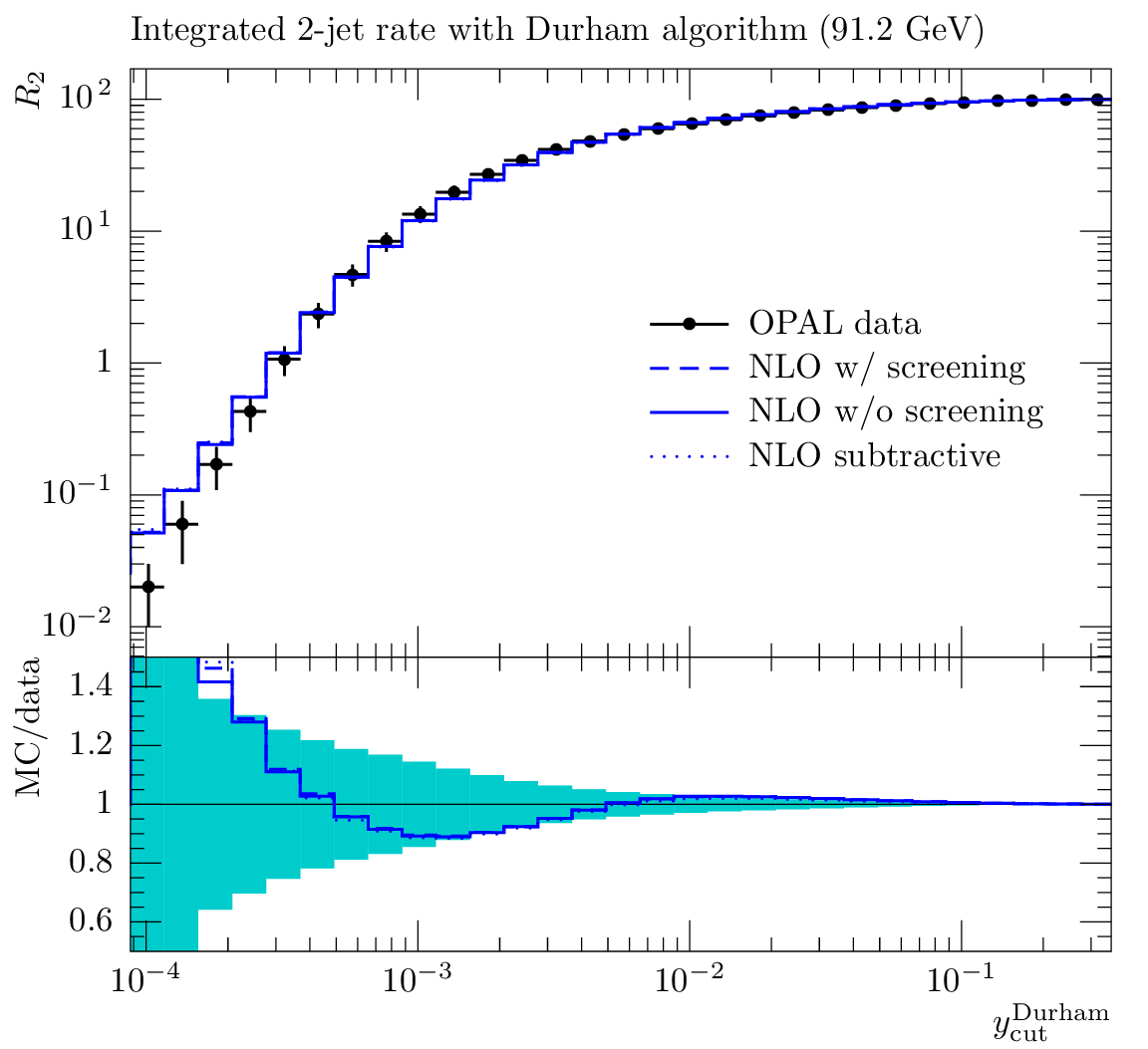}
\caption{\label{figures:lepeec-comparematchings}Comparison of matching strategies
exemplified for the Durham two-jet rate.}
\end{figure}

The \matchbox{} framework provides the facility to switch between the
\powheg{}-type matching with matrix element corrections including or
excluding the auxiliary Born screening cross section, and subtractive
matching.  For reasons of systematics it is instructive to compare
these approaches.  No separate fit for the variants not considered so
far has been performed and the NLO fit values as given in the previous
section have been used.  The different matching strategies give
completely comparable results.  If there are small visible
differences, there is no clear tendency that either variant would give
a better description than any of the others.
Fig.~\ref{figures:lepeec-comparematchings} compares the matching
strategies for the two jet rate. To this extent, the subtractive
matching could be preferred amongst the \powheg-type ones owing to its
smaller computational complexity. This statement, however, not only
includes that negative weighted events do not pose a major problem,
but also has to be verified in a process dependent matter since there
is no hint, if the behaviour observed here is a general feature --
particularly at hadron colliders.

\section{Results at HERA}
\label{sections:hera}

Owing to the approximation underlying the dipole parton shower,
diagrams contributing to parton emission of a given dipole $(i,j)$ may
be considered a gauge invariant subset in the soft and/or collinear
limits for $N_c\to\infty$.  This implies that the infrared cutoffs
and soft scales entering the emission probabilities need not be the
same for all dipoles. The emitter-spectator configurations forming
gauge invariant quantities in this sense are the two emitter choices
for final-final dipoles, initial-initial dipoles, and the combination
of initial-final and final-initial configurations.  Fitting DIS data
therefore allows one to fix the infrared cutoff and soft scale for the
latter, before finally constraining the same parameters for
initial-initial dipoles at a hadron collider, which is considered in
the next section.

For the fit described here, the same technique as for LEP, and data
accumulated by the H1 experiment \cite{Adloff:1999ws} have been
used. For LO and NLO, the default \hpp{} PDFs, MSTW 2008 LO**
\cite{Martin:2009iq,Sherstnev:2007nd} and MRST 2002 NLO
\cite{Thorne:2002mr}, have been used. The same PDFs were considered
for hadron collider data to be discussed in the next section. The NLO
fit was obtained by running the matching with matrix element
correction.

The findings are similar as for the fit to LEP data. We find a
reasonable prediction of transverse energy flows over the whole range
of $(x,Q^2)$ plane.  The matched NLO prediction gives a comparable fit
to the LO simulation, while preferring both a smaller infrared cutoff
and screening scale. The fitted parameters are given in
Tab.~\ref{tables:disfittedparameters}.

\begin{figure}
\begin{center}
\includegraphics[scale=0.6]{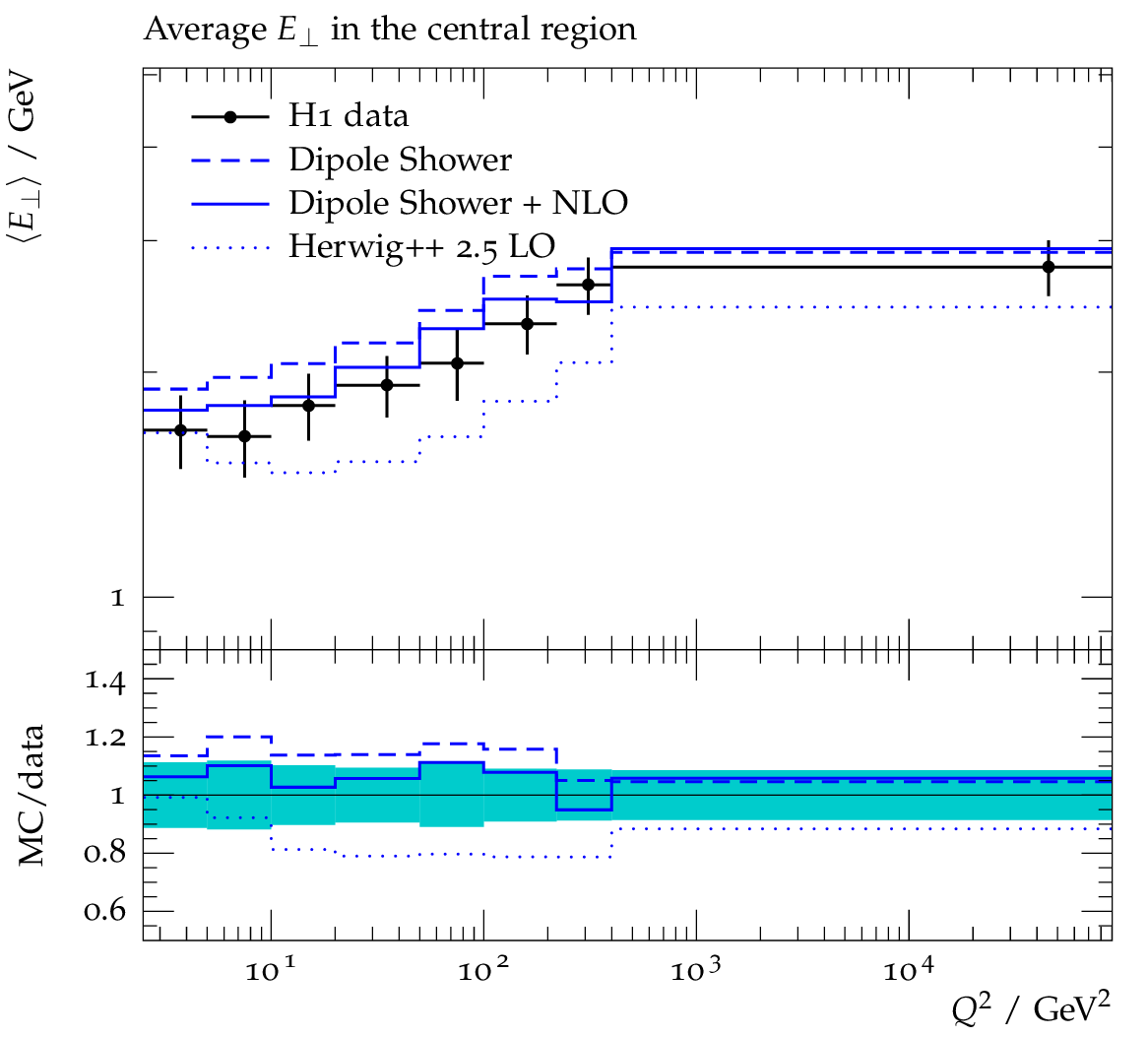}
\end{center}
\caption{\label{figures:H1transverse}Average transverse energy 
in the central region as measured at HERA and
compared to leading order and next-to-leading order predictions.}
\end{figure}

Fig.~\ref{figures:H1transverse} shows the average transverse energy
as a function of $Q^2$ in the central detector
region.  This observable is clearly improved by the NLO matching
at small momentum transfers. A more detailed analysis of DIS data
including inclusive jet and event shape data is currently underway.

\section{Results at the Tevatron}
\label{sections:tev}

After having determined the simulation parameters for hadronization,
final state radiation, and radiation off a final-initial dipole
by fitting LEP and HERA data, two parameters remain to be determined:
the infrared cutoff and soft scale
for radiation off an initial-initial dipole. We here consider the $p_\perp$
spectrum of $e^+e^-$ Drell-Yan pair production as measured by the
CDF collaboration \cite{Affolder:1999jh}. Since the Drell-Yan
process receives rather large QCD corrections from 
leading to next-to-leading order and a still considerable correction
at NNLO, both fits have been performed by normalising the simulation
to the measured cross section. The matrix element matching including
the Born screening cross section has been used here, as for the
DIS data.

The \professor{} algorithm here turned out not to be applicable, as the
cubic interpolation was not capable of describing the complete
dynamics of letting the shower evolve to rather small infrared
cutoffs, owing to the prescription of introducing a soft scale in
$\alpha_s$ as already described before. We have therefore performed a
preliminary fit by generating $300$ random points uniformly in
parameter space, which here includes the infrared cutoff for
initial-initial dipoles, the soft scale for initial-initial dipoles,
as well as the widths of a Gaussian distribution for intrinsic
transverse momentum, $\Lambda_\perp$. The latter has been chosen to be
potentially different for valence and sea partons.

Out of these random points we have picked the one with lowest
$\chi^2$ with respect to the data, again both for LO and NLO
simulations. The resulting parameters are given in
Tab.~\ref{tables:tvtfittedparameters}. Note that the $p_\perp$
distribution for sea partons is narrower, corresponding to a broader
spatial distribution as can be motivated on different grounds.  

\begin{figure}
\begin{center}
\includegraphics[scale=0.6]{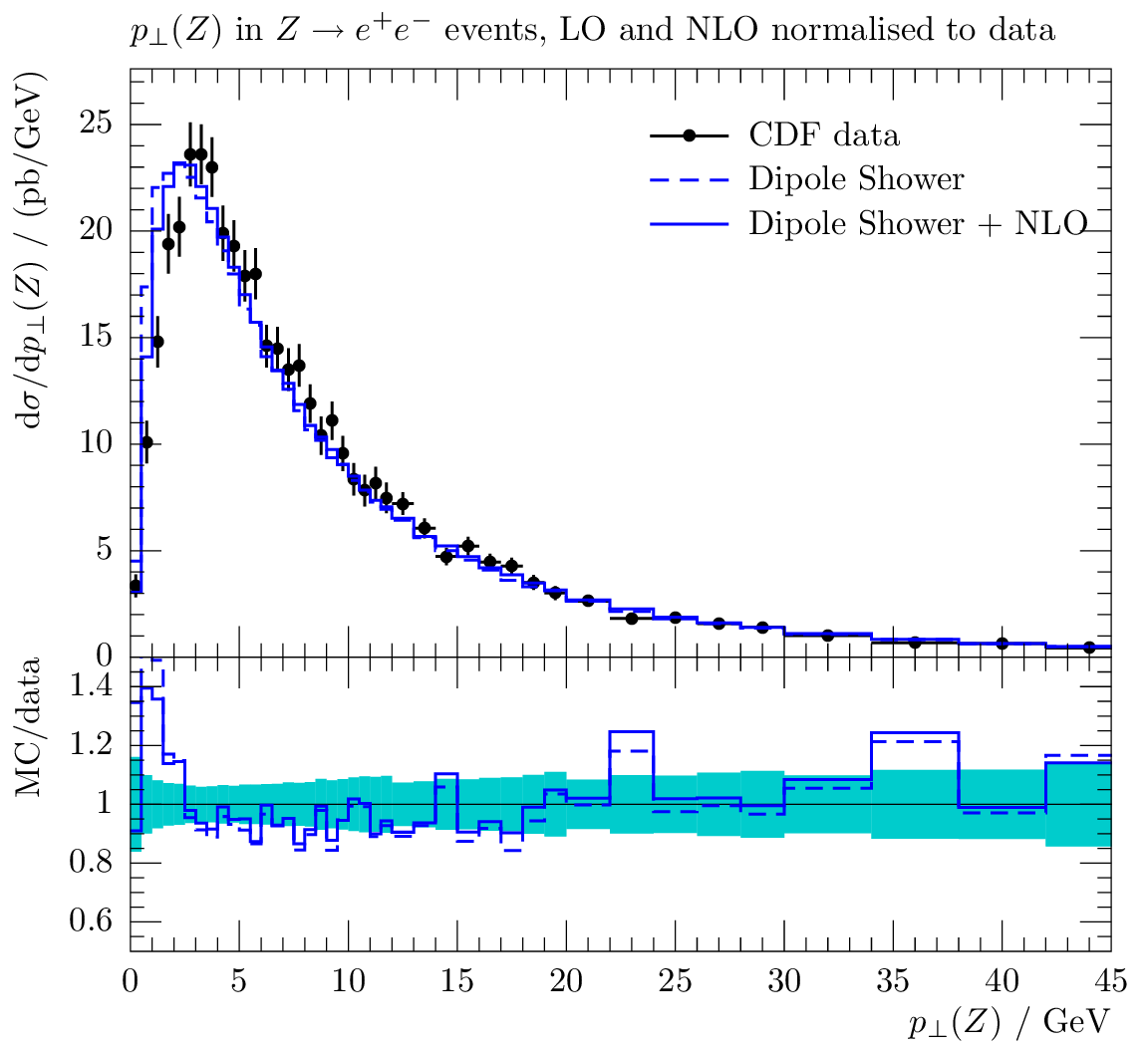}
\end{center}
\caption{\label{figures:tvtdy}Differential cross section of the
  Drell-Yan-pair $p_\perp$ compared to LO and NLO predictions. Note
  that the cross sections have been normalised to the measured one.}
\end{figure}

\begin{figure}
\begin{center}
\includegraphics[scale=0.6]{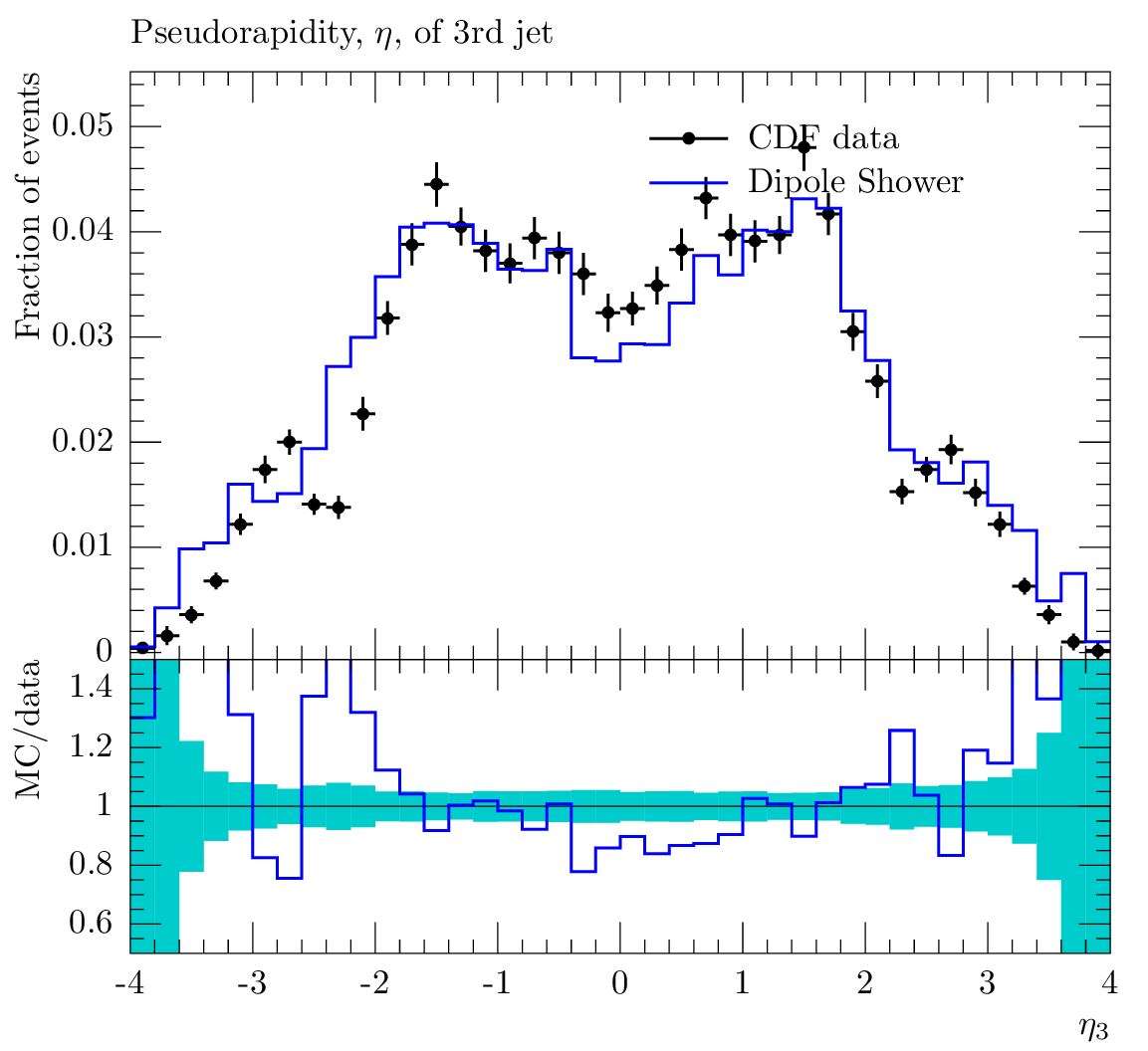}
\end{center}
\caption{\label{figures:cdfjets} The pseudo-rapidity distribution of a
  third jet in events with at least two jets. We here only show the
  leading order prediction in order to check the predictivity of the
  tune carried out so far.}
\end{figure}

We show the comparison of LO and NLO simulations in
Fig.~\ref{figures:tvtdy} showing similar systematics to the
distributions discussed before. In order to determine the predictivity
of the simulation already at this very coarse level of tuning, we
additionally show the pseudo-rapidity distribution of a third jet in
events with at least two hard jets, Fig.~\ref{figures:cdfjets}, as
carried out at CDF \cite{Abe:1994nj}. Reasonable agreement with data
is found. On top of the work presented in \cite{Platzer:2009jq}, this
constitutes another crucial test of coherent parton evolution.

\section{Conclusions}
\label{sections:conclusions}
We have introduced a new dipole shower module for the event generator
\hpp{} that allows for an automatic matching of NLO computations with a
parton shower.  A tune of the hadronization module to the most important
data sets show that we can achieve very good results from this
simulation already without the inclusion of NLO terms.   Including NLO
corrections at this relatively simple level only marginally improves the
results.  This effect is expected as it is known that the
Catani--Seymour showers tend to mimic the behaviour of NLO matrix
elements very well also in phase space regions well outside the
collinear limits.   However, the matching poses no technical problem and
can be seen as a proof--of--concept for the idea to provide a framework
for automatic matching.  At this time with relatively simple matrix
elements at NLO that are provided by internal code. Future work
will concentrate in the inclusion of external code via a well defined
interface, following the ideas in \cite{Binoth:2010xt}.  

\section*{Acknowledgements}
We are grateful to the other \hpp{} authors and Leif L\"onnblad for
extensive collaboration. We would like to thank Keith Hamilton,
Christian R\"ohr and Mike Seymour for extensive comments on the
manuscript.  This work was supported in part by the European Union
Marie Curie Research Training Network MCnet under contract
MRTN-CT-2006-035606 and the Helmholtz Alliance ``Physics at the
Terascale''.

\bibliography{matchbox}

\appendix

\section{Code Validation}
\label{sections:validation}

\subsection{Shower Splitting Kernels}

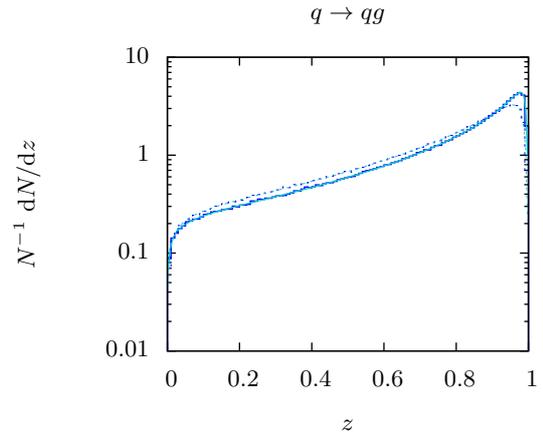
\begin{figure}
\centering
\input{ff_qx2qgx_z}
\caption{\label{figures:kernelslep}Example comparison of sampled
  final-final splitting momentum fraction (blue lines) versus results
  from a numerical integration (turquoise lines) at two different
  dipole masses, $s_{ij}=(100 {\rm GeV})^2$ (continuous lines) and
  $s_{ij}=(50 {\rm GeV})^2$ (broken lines).}
\end{figure}

The sampling of shower splitting kernels has been explicitly verified
{\it in situ}, meaning using the full implementation as present in the
simulation code, against an independent implementation using a
numerical integration to obtain the Sudakov-type
distributions. Fig.~\ref{figures:kernelslep} shows an example for a
final-final splitting kernel, proving correctness of this part of
the code.

\subsection{NLO QCD Corrections}

All leading order matrix elements implemented in the
\matchbox{} framework have been cross-checked against
the \hpp{} matrix elements.

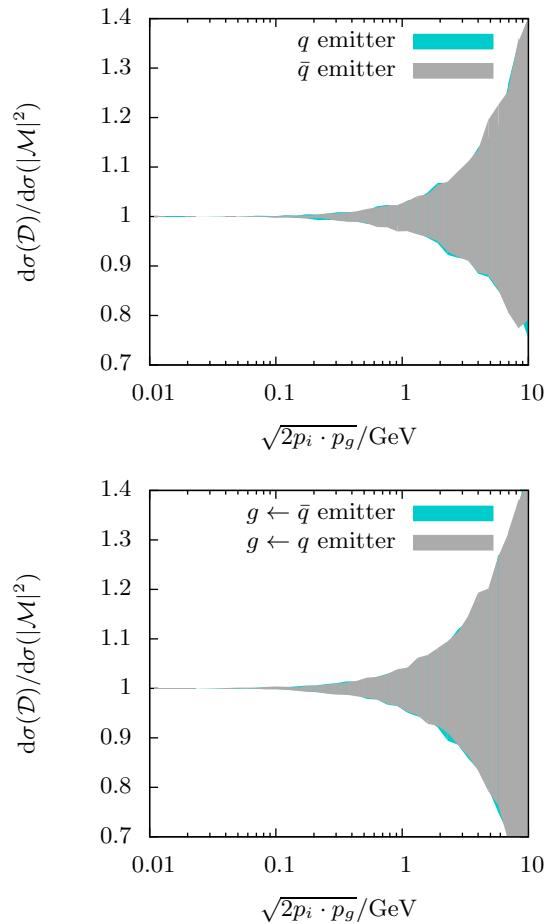
\begin{figure}
\centering
\input{sub-tvt-qqbar} \input{sub-tvt-gq}
\caption{\label{figures:subtvt} Envelopes of the ratio of the subtraction
to the real emission cross section versus the propagator denominator for
all singular configurations in $Z+{\rm jet}$ production.}
\end{figure}

The functionality of the automatically generated subtraction terms has
been verified.  Fig.~\ref{figures:subtvt} shows a typical examples
of the ratio of subtraction to real emission cross section, plotted
against each of the invariants entering the propagator denominators.

The `plain' NLO cross section, and the inclusive one entering the
matching with matrix element correction have been checked to agree,
with and without the usage of the Born `screening' cross section.  The
NLO cross section for $e^+e^-\to {\rm jets}$ has been validated
against the analytically known $K$-factor of $1+\alpha_s/\pi$.  The
NLO cross section for DIS and Drell-Yan has been checked against the
existing \powheg{} implementation in \hpp{}.  For deep
inelastic scattering, the subtraction terms have been modified in
order to have positive definite dipole kernels, finite terms of the
integrated subtraction terms have been changed accordingly. The
functionality of the subtraction has been checked with both variants,
and the NLO cross sections with and without modifications are found to
agree.

\subsection{NLO Matching with Matrix Element Corrections}

A non-trivial cross check of the matrix element correction code and
\exsample{} as the underlying `working horse', is to consider
the spectra for a gluon emission off a $q\bar{q}$ dipole as generated
by the shower, which is validated against a numerical integration of
the expected distribution implemented in a completely independent
code. By putting the real emission matrix element entering the
matching to be equal to the sum of dipoles (the correctness of which
has been checked by verifying that the cross section of the subtracted
real emission matrix element is consistent with zero), the matrix
element correction must produce the same spectrum as the shower
code. We have checked that this is indeed the case. It should be
stressed that the machinery underlying the setup of the matrix element
correction is much more complex than the shower implementation, and,
that the splitting kernel entering the matrix element correction does
depend on more parameters\footnote{ In a realistic application these
  are not two random numbers needed for the Born process, but indeed
  six, since photon radiation is generated of each incoming lepton,
  requiring two random numbers per incoming lepton.}  than the one
parameter of the shower kernel (corresponding to the dipole invariant
mass).

\onecolumn

\begin{table}
\centering
\begin{tabular}{|p{2cm}|p{4cm}|p{6.5cm}|}
\hline
{\bf Parameter} & {\bf Range} & {\bf Description} \\ \hline
$\alpha_s(M_Z^2)$ & $0.1 - 0.13$ & Input $\alpha_s$ at $Z$ mass. \\ \hline
$\mu_{IR,FF}$ & $0.5\ {\rm GeV} - 2.0\ {\rm GeV}$ & Infrared cutoff for final-final dipoles\\ \hline
$\mu_{\text{soft},FF}$ & $0.0\ {\rm GeV} - 1.2\ {\rm GeV}$ & Soft scale for final-final dipoles\\ \hline
$m_{g,c}$ & $0.67 \ {\rm GeV} - 3.0 \ {\rm GeV}$& Gluon constituent mass\\ \hline
$\text{Cl}_{\text{max}}$ & $0.5\ {\rm GeV} - 10\ {\rm GeV}$ & Maximum cluster mass\\ \hline
$\text{Cl}_{\text{pow}}$ & $0.0 - 10.0$ & Cluster mass exponent\\ \hline
$\text{Cl}_{\text{smr}}$ & $0.0 - 10.0$ & Cluster direction smearing\\ \hline
$P_{\text{split}}$ & $0.0 - 1.4$ & Cluster mass splitting parameter\\ \hline
\end{tabular}
\caption{\label{tables:lepfitparameters}The parameters varied for
the fit to LEP data.}
\end{table}

\begin{table}
\centering
\begin{tabular}{|p{2cm}|p{5cm}|p{5cm}|}
\hline
{\bf Parameter} & {\bf LO} & {\bf NLO} \\ \hline
$\alpha_s(M_Z^2)$ & $0.113185 \pm 0.007281$ & $ 0.117550\pm 0.005053$ \\ \hline
$\mu_{IR,FF}$ & $(1.416023 \pm 0.306430)\ {\rm GeV}$ & $(1.245196 \pm 0.226821)\ {\rm GeV}$ \\ \hline
$\mu_{\text{soft},FF}$ & $(0.242725 \pm 0.202069 )\ {\rm GeV}$ & $0.0\ {\rm GeV}$ \footnotemark \\ \hline
$m_{g,c}$ & $(1.080386 \pm 0.499546)\ {\rm GeV}$ & $(1.007680 \pm 0.265565 )\ {\rm GeV}$ \\ \hline
$\text{Cl}_{\text{max}}$ & $(4.170320 \pm 0.589504)\ {\rm GeV}$ & $(3.664004 \pm 0.639504)\ {\rm GeV}$ \\ \hline
$\text{Cl}_{\text{pow}}$ & $5.734681 \pm 1.006965 $ & $5.687022 \pm 0.869322$ \\ \hline
$\text{Cl}_{\text{smr}}$ & $4.548755 \pm 2.350193 $ & $3.115744 \pm 2.436793$ \\ \hline
$P_{\text{split}}$ & $0.765173 \pm 0.074008$ & $0.771329 \pm 0.074248$ \\ \hline
\end{tabular}
\caption{\label{tables:lepfittedparameters}Parameters for LO and NLO fits to LEP data.}
\end{table}
\footnotetext{This parameter was predicted negative by \professor{} though
consistent with zero and has thus been fixed.}

\begin{table}
\centering
\begin{tabular}{|p{2cm}|p{5cm}|p{5cm}|}
\hline
{\bf Parameter} & {\bf LO} & {\bf NLO} \\ \hline
$\mu_{IR,FI}$ & $(0.796205 \pm 0.333340 )\ {\rm GeV}$ & $(0.718418 \pm 0.210448 )\ {\rm GeV}$ \\ \hline
$\mu_{\text{soft},FI}$ & $(1.355894 \pm 0.432515 )\ {\rm GeV}$ & $( 1.003714\pm 0.252398  ) {\rm GeV}$ \\ \hline
\end{tabular}
\caption{\label{tables:disfittedparameters}Parameters for LO and NLO fits to HERA data.}
\end{table}

\begin{table}
\centering
\begin{tabular}{|p{2cm}|p{5cm}|p{5cm}|}
\hline
{\bf Parameter} & {\bf LO} & {\bf NLO} \\ \hline
$\mu_{IR,II}$ & $0.367359\ {\rm GeV}$ & $0.275894\ {\rm GeV}$ \\ \hline
$\mu_{\text{soft},II}$ & $0.205854\ {\rm GeV}$ & $0.254028\ {\rm GeV}$ \\ \hline
$\Lambda_{\perp,\text{valence}}$ & $1.68463\ {\rm GeV}$ & $1.26905\ {\rm GeV}$ \\ \hline
$\Lambda_{\perp,\text{sea}}$ & $1.29001\ {\rm GeV}$ & $1.1613\ {\rm GeV}$ \\ \hline
\end{tabular}
\caption{\label{tables:tvtfittedparameters}Parameters for LO and NLO
  fits to the CDF Drell-Yan data.}
\end{table}

\twocolumn

\end{document}

%% file: ff_qx2qgx_z.tex
\begingroup
  \makeatletter
  \providecommand\color[2][]{%
    \GenericError{(gnuplot) \space\space\space\@spaces}{%
      Package color not loaded in conjunction with
      terminal option `colourtext'%
    }{See the gnuplot documentation for explanation.%
    }{Either use 'blacktext' in gnuplot or load the package
      color.sty in LaTeX.}%
    \renewcommand\color[2][]{}%
  }%
  \providecommand\includegraphics[2][]{%
    \GenericError{(gnuplot) \space\space\space\@spaces}{%
      Package graphicx or graphics not loaded%
    }{See the gnuplot documentation for explanation.%
    }{The gnuplot epslatex terminal needs graphicx.sty or graphics.sty.}%
    \renewcommand\includegraphics[2][]{}%
  }%
  \providecommand\rotatebox[2]{#2}%
  \@ifundefined{ifGPcolor}{%
    \newif\ifGPcolor
    \GPcolortrue
  }{}%
  \@ifundefined{ifGPblacktext}{%
    \newif\ifGPblacktext
    \GPblacktexttrue
  }{}%
  \let\gplgaddtomacro\g@addto@macro
  \gdef\gplbacktext{}%
  \gdef\gplfronttext{}%
  \makeatother
  \ifGPblacktext
    \def\colorrgb#1{}%
    \def\colorgray#1{}%
  \else
    \ifGPcolor
      \def\colorrgb#1{\color[rgb]{#1}}%
      \def\colorgray#1{\color[gray]{#1}}%
      \expandafter\def\csname LTw\endcsname{\color{white}}%
      \expandafter\def\csname LTb\endcsname{\color{black}}%
      \expandafter\def\csname LTa\endcsname{\color{black}}%
      \expandafter\def\csname LT0\endcsname{\color[rgb]{1,0,0}}%
      \expandafter\def\csname LT1\endcsname{\color[rgb]{0,1,0}}%
      \expandafter\def\csname LT2\endcsname{\color[rgb]{0,0,1}}%
      \expandafter\def\csname LT3\endcsname{\color[rgb]{1,0,1}}%
      \expandafter\def\csname LT4\endcsname{\color[rgb]{0,1,1}}%
      \expandafter\def\csname LT5\endcsname{\color[rgb]{1,1,0}}%
      \expandafter\def\csname LT6\endcsname{\color[rgb]{0,0,0}}%
      \expandafter\def\csname LT7\endcsname{\color[rgb]{1,0.3,0}}%
      \expandafter\def\csname LT8\endcsname{\color[rgb]{0.5,0.5,0.5}}%
    \else
      \def\colorrgb#1{\color{black}}%
      \def\colorgray#1{\color[gray]{#1}}%
      \expandafter\def\csname LTw\endcsname{\color{white}}%
      \expandafter\def\csname LTb\endcsname{\color{black}}%
      \expandafter\def\csname LTa\endcsname{\color{black}}%
      \expandafter\def\csname LT0\endcsname{\color{black}}%
      \expandafter\def\csname LT1\endcsname{\color{black}}%
      \expandafter\def\csname LT2\endcsname{\color{black}}%
      \expandafter\def\csname LT3\endcsname{\color{black}}%
      \expandafter\def\csname LT4\endcsname{\color{black}}%
      \expandafter\def\csname LT5\endcsname{\color{black}}%
      \expandafter\def\csname LT6\endcsname{\color{black}}%
      \expandafter\def\csname LT7\endcsname{\color{black}}%
      \expandafter\def\csname LT8\endcsname{\color{black}}%
    \fi
  \fi
  \setlength{\unitlength}{0.0500bp}%
  \begin{picture}(4320.00,3528.00)%
    \gplgaddtomacro\gplbacktext{%
      \csname LTb\endcsname%
      \put(1122,660){\makebox(0,0)[r]{\strut{} 0.01}}%
      \put(1122,1396){\makebox(0,0)[r]{\strut{} 0.1}}%
      \put(1122,2132){\makebox(0,0)[r]{\strut{} 1}}%
      \put(1122,2868){\makebox(0,0)[r]{\strut{} 10}}%
      \put(1254,440){\makebox(0,0){\strut{} 0}}%
      \put(1792,440){\makebox(0,0){\strut{} 0.2}}%
      \put(2331,440){\makebox(0,0){\strut{} 0.4}}%
      \put(2869,440){\makebox(0,0){\strut{} 0.6}}%
      \put(3408,440){\makebox(0,0){\strut{} 0.8}}%
      \put(3946,440){\makebox(0,0){\strut{} 1}}%
      \put(220,1764){\rotatebox{90}{\makebox(0,0){\strut{}$N^{-1}\ {\rm d}N/ {\rm d}z$}}}%
      \put(2600,110){\makebox(0,0){\strut{}$z$}}%
      \put(2600,3198){\makebox(0,0){\strut{}$q\to q g$}}%
    }%
    \gplgaddtomacro\gplfronttext{%
    }%
    \gplbacktext
    \put(0,0){\includegraphics{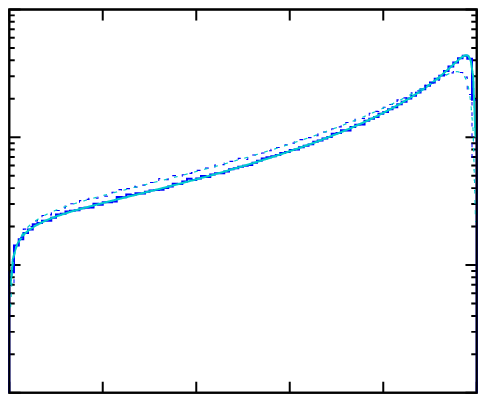}}%
    \gplfronttext
  \end{picture}%
\endgroup

%% file: sub-tvt-qqbar.tex
\begingroup
  \makeatletter
  \providecommand\color[2][]{%
    \GenericError{(gnuplot) \space\space\space\@spaces}{%
      Package color not loaded in conjunction with
      terminal option `colourtext'%
    }{See the gnuplot documentation for explanation.%
    }{Either use 'blacktext' in gnuplot or load the package
      color.sty in LaTeX.}%
    \renewcommand\color[2][]{}%
  }%
  \providecommand\includegraphics[2][]{%
    \GenericError{(gnuplot) \space\space\space\@spaces}{%
      Package graphicx or graphics not loaded%
    }{See the gnuplot documentation for explanation.%
    }{The gnuplot epslatex terminal needs graphicx.sty or graphics.sty.}%
    \renewcommand\includegraphics[2][]{}%
  }%
  \providecommand\rotatebox[2]{#2}%
  \@ifundefined{ifGPcolor}{%
    \newif\ifGPcolor
    \GPcolortrue
  }{}%
  \@ifundefined{ifGPblacktext}{%
    \newif\ifGPblacktext
    \GPblacktexttrue
  }{}%
  \let\gplgaddtomacro\g@addto@macro
  \gdef\gplbacktext{}%
  \gdef\gplfronttext{}%
  \makeatother
  \ifGPblacktext
    \def\colorrgb#1{}%
    \def\colorgray#1{}%
  \else
    \ifGPcolor
      \def\colorrgb#1{\color[rgb]{#1}}%
      \def\colorgray#1{\color[gray]{#1}}%
      \expandafter\def\csname LTw\endcsname{\color{white}}%
      \expandafter\def\csname LTb\endcsname{\color{black}}%
      \expandafter\def\csname LTa\endcsname{\color{black}}%
      \expandafter\def\csname LT0\endcsname{\color[rgb]{1,0,0}}%
      \expandafter\def\csname LT1\endcsname{\color[rgb]{0,1,0}}%
      \expandafter\def\csname LT2\endcsname{\color[rgb]{0,0,1}}%
      \expandafter\def\csname LT3\endcsname{\color[rgb]{1,0,1}}%
      \expandafter\def\csname LT4\endcsname{\color[rgb]{0,1,1}}%
      \expandafter\def\csname LT5\endcsname{\color[rgb]{1,1,0}}%
      \expandafter\def\csname LT6\endcsname{\color[rgb]{0,0,0}}%
      \expandafter\def\csname LT7\endcsname{\color[rgb]{1,0.3,0}}%
      \expandafter\def\csname LT8\endcsname{\color[rgb]{0.5,0.5,0.5}}%
    \else
      \def\colorrgb#1{\color{black}}%
      \def\colorgray#1{\color[gray]{#1}}%
      \expandafter\def\csname LTw\endcsname{\color{white}}%
      \expandafter\def\csname LTb\endcsname{\color{black}}%
      \expandafter\def\csname LTa\endcsname{\color{black}}%
      \expandafter\def\csname LT0\endcsname{\color{black}}%
      \expandafter\def\csname LT1\endcsname{\color{black}}%
      \expandafter\def\csname LT2\endcsname{\color{black}}%
      \expandafter\def\csname LT3\endcsname{\color{black}}%
      \expandafter\def\csname LT4\endcsname{\color{black}}%
      \expandafter\def\csname LT5\endcsname{\color{black}}%
      \expandafter\def\csname LT6\endcsname{\color{black}}%
      \expandafter\def\csname LT7\endcsname{\color{black}}%
      \expandafter\def\csname LT8\endcsname{\color{black}}%
    \fi
  \fi
  \setlength{\unitlength}{0.0500bp}%
  \begin{picture}(4320.00,3528.00)%
    \gplgaddtomacro\gplbacktext{%
      \csname LTb\endcsname%
      \put(990,660){\makebox(0,0)[r]{\strut{} 0.7}}%
      \put(990,1032){\makebox(0,0)[r]{\strut{} 0.8}}%
      \put(990,1404){\makebox(0,0)[r]{\strut{} 0.9}}%
      \put(990,1776){\makebox(0,0)[r]{\strut{} 1}}%
      \put(990,2148){\makebox(0,0)[r]{\strut{} 1.1}}%
      \put(990,2520){\makebox(0,0)[r]{\strut{} 1.2}}%
      \put(990,2892){\makebox(0,0)[r]{\strut{} 1.3}}%
      \put(990,3264){\makebox(0,0)[r]{\strut{} 1.4}}%
      \put(1122,440){\makebox(0,0){\strut{} 0.01}}%
      \put(2063,440){\makebox(0,0){\strut{} 0.1}}%
      \put(3005,440){\makebox(0,0){\strut{} 1}}%
      \put(3946,440){\makebox(0,0){\strut{} 10}}%
      \put(220,1962){\rotatebox{90}{\makebox(0,0){\strut{}${\rm d}\sigma({\cal D})/{\rm d}\sigma(|{\cal M}|^2)$}}}%
      \put(2534,110){\makebox(0,0){\strut{}$\sqrt{2 p_i\cdot p_g}$/{\rm GeV}}}%
    }%
    \gplgaddtomacro\gplfronttext{%
      \csname LTb\endcsname%
      \put(2959,3091){\makebox(0,0)[r]{\strut{}$q$ emitter}}%
      \csname LTb\endcsname%
      \put(2959,2871){\makebox(0,0)[r]{\strut{}$\bar{q}$ emitter}}%
    }%
    \gplbacktext
    \put(0,0){\includegraphics{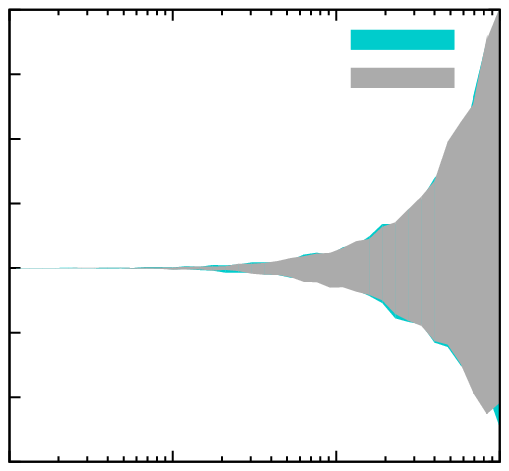}}%
    \gplfronttext
  \end{picture}%
\endgroup

%% file: sub-tvt-gq.tex
\begingroup
  \makeatletter
  \providecommand\color[2][]{%
    \GenericError{(gnuplot) \space\space\space\@spaces}{%
      Package color not loaded in conjunction with
      terminal option `colourtext'%
    }{See the gnuplot documentation for explanation.%
    }{Either use 'blacktext' in gnuplot or load the package
      color.sty in LaTeX.}%
    \renewcommand\color[2][]{}%
  }%
  \providecommand\includegraphics[2][]{%
    \GenericError{(gnuplot) \space\space\space\@spaces}{%
      Package graphicx or graphics not loaded%
    }{See the gnuplot documentation for explanation.%
    }{The gnuplot epslatex terminal needs graphicx.sty or graphics.sty.}%
    \renewcommand\includegraphics[2][]{}%
  }%
  \providecommand\rotatebox[2]{#2}%
  \@ifundefined{ifGPcolor}{%
    \newif\ifGPcolor
    \GPcolortrue
  }{}%
  \@ifundefined{ifGPblacktext}{%
    \newif\ifGPblacktext
    \GPblacktexttrue
  }{}%
  \let\gplgaddtomacro\g@addto@macro
  \gdef\gplbacktext{}%
  \gdef\gplfronttext{}%
  \makeatother
  \ifGPblacktext
    \def\colorrgb#1{}%
    \def\colorgray#1{}%
  \else
    \ifGPcolor
      \def\colorrgb#1{\color[rgb]{#1}}%
      \def\colorgray#1{\color[gray]{#1}}%
      \expandafter\def\csname LTw\endcsname{\color{white}}%
      \expandafter\def\csname LTb\endcsname{\color{black}}%
      \expandafter\def\csname LTa\endcsname{\color{black}}%
      \expandafter\def\csname LT0\endcsname{\color[rgb]{1,0,0}}%
      \expandafter\def\csname LT1\endcsname{\color[rgb]{0,1,0}}%
      \expandafter\def\csname LT2\endcsname{\color[rgb]{0,0,1}}%
      \expandafter\def\csname LT3\endcsname{\color[rgb]{1,0,1}}%
      \expandafter\def\csname LT4\endcsname{\color[rgb]{0,1,1}}%
      \expandafter\def\csname LT5\endcsname{\color[rgb]{1,1,0}}%
      \expandafter\def\csname LT6\endcsname{\color[rgb]{0,0,0}}%
      \expandafter\def\csname LT7\endcsname{\color[rgb]{1,0.3,0}}%
      \expandafter\def\csname LT8\endcsname{\color[rgb]{0.5,0.5,0.5}}%
    \else
      \def\colorrgb#1{\color{black}}%
      \def\colorgray#1{\color[gray]{#1}}%
      \expandafter\def\csname LTw\endcsname{\color{white}}%
      \expandafter\def\csname LTb\endcsname{\color{black}}%
      \expandafter\def\csname LTa\endcsname{\color{black}}%
      \expandafter\def\csname LT0\endcsname{\color{black}}%
      \expandafter\def\csname LT1\endcsname{\color{black}}%
      \expandafter\def\csname LT2\endcsname{\color{black}}%
      \expandafter\def\csname LT3\endcsname{\color{black}}%
      \expandafter\def\csname LT4\endcsname{\color{black}}%
      \expandafter\def\csname LT5\endcsname{\color{black}}%
      \expandafter\def\csname LT6\endcsname{\color{black}}%
      \expandafter\def\csname LT7\endcsname{\color{black}}%
      \expandafter\def\csname LT8\endcsname{\color{black}}%
    \fi
  \fi
  \setlength{\unitlength}{0.0500bp}%
  \begin{picture}(4320.00,3528.00)%
    \gplgaddtomacro\gplbacktext{%
      \csname LTb\endcsname%
      \put(990,660){\makebox(0,0)[r]{\strut{} 0.7}}%
      \put(990,1032){\makebox(0,0)[r]{\strut{} 0.8}}%
      \put(990,1404){\makebox(0,0)[r]{\strut{} 0.9}}%
      \put(990,1776){\makebox(0,0)[r]{\strut{} 1}}%
      \put(990,2148){\makebox(0,0)[r]{\strut{} 1.1}}%
      \put(990,2520){\makebox(0,0)[r]{\strut{} 1.2}}%
      \put(990,2892){\makebox(0,0)[r]{\strut{} 1.3}}%
      \put(990,3264){\makebox(0,0)[r]{\strut{} 1.4}}%
      \put(1122,440){\makebox(0,0){\strut{} 0.01}}%
      \put(2063,440){\makebox(0,0){\strut{} 0.1}}%
      \put(3005,440){\makebox(0,0){\strut{} 1}}%
      \put(3946,440){\makebox(0,0){\strut{} 10}}%
      \put(220,1962){\rotatebox{90}{\makebox(0,0){\strut{}${\rm d}\sigma({\cal D})/{\rm d}\sigma(|{\cal M}|^2)$}}}%
      \put(2534,110){\makebox(0,0){\strut{}$\sqrt{2 p_i\cdot p_g}$/{\rm GeV}}}%
    }%
    \gplgaddtomacro\gplfronttext{%
      \csname LTb\endcsname%
      \put(2959,3091){\makebox(0,0)[r]{\strut{}$g\gets \bar{q}$ emitter}}%
      \csname LTb\endcsname%
      \put(2959,2871){\makebox(0,0)[r]{\strut{}$g\gets q$ emitter}}%
    }%
    \gplbacktext
    \put(0,0){\includegraphics{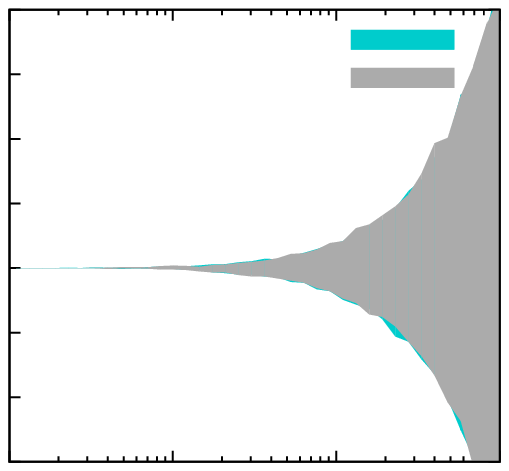}}%
    \gplfronttext
  \end{picture}%
\endgroup